\documentclass[twocolumn,showpacs,aps,prl,superscriptaddress,floatfix,nobibnotes]{revtex4}

\usepackage{epsfig}
\usepackage[hang,center]{subfigure}
\usepackage{multirow,tabls,amssymb,amsmath}

\input pubboard/babarsym

\newcommand{\Lambdac}           {\ensuremath{\Lambda_{c}^{+}}\xspace}
\newcommand{\Sigmac}            {\ensuremath{\Sigma_{c}}\xspace}
\newcommand{\Sigcz}             {\ensuremath{\Sigma_{c}^{0}}\xspace}
\newcommand{\Sigcp}             {\ensuremath{\Sigma_{c}^{+}}\xspace}

\newcommand{\Sigmacz}           {\ensuremath{\Sigma_{c}(2455)^{0}}\xspace}

\newcommand{\Sigmaczstar}       {\ensuremath{\Sigma_{c}(2520)^{0}}\xspace}

\newcommand{\Sigmaczstarstar}   {\ensuremath{\Sigma_{c}(2800)^{0}}\xspace}
\newcommand{\Sigmacpstarstar}   {\ensuremath{\Sigma_{c}(2800)^{+}}\xspace}
\newcommand{\Sigmacppstarstar}   {\ensuremath{\Sigma_{c}(2800)^{++}}\xspace}
\newcommand{\SigctoLcpi}        {\ensuremath{\Sigmacz \to \Lambdac \pim}\xspace}

\newcommand{\BzbtoLcpb}         {\ensuremath{\Bzb \to \Lambdac \antiproton}\xspace}

\newcommand{\BmtoLcpbpi}        {\ensuremath{\Bm \to \Lambdac \antiproton \pim}\xspace}

\newcommand{\BmtoSigcpb}        {\ensuremath{\Bm \to \Sigmacz \antiproton}\xspace}
\newcommand{\BmtoSigcstarpb}    {\ensuremath{\Bm \to \Sigma_{c}(2520)^{0} \antiproton}\xspace}
\newcommand{\BmtoSigcstarstarpb}    {\ensuremath{\Bm \to \Sigma_{c}(2800)^{0} \antiproton}\xspace}
\newcommand{\BmtoLambdaDelta}   {\ensuremath{\Bm \to \Lambdac \bar{\Delta}(1232)^{--}}\xspace}

\newcommand{\pKpi}              {\ensuremath{\proton \Km \pip}\xspace}
\newcommand{\pKs}               {\ensuremath{\proton \KS}\xspace}
\newcommand{\pKspipi}           {\ensuremath{\proton \KS \pip \pim}\xspace}
\newcommand{\Lambdapi}          {\ensuremath{\Lambda \pip}\xspace}
\newcommand{\Lambdapipipi}      {\ensuremath{\Lambda \pip \pim \pip}\xspace}
\newcommand{\Lcp}               {\ensuremath{\Lambdac \antiproton}\xspace}
\newcommand{\Lcpi}               {\ensuremath{\Lambdac \pim}\xspace}
\newcommand{\splot}             {\ensuremath{_{s}\mathcal{P}lot}\xspace}

\newcommand{\LctopKpi}          {\ensuremath{\Lambdac \to \proton \Km \pip}\xspace}

\newcommand{\LctopKs}           {\ensuremath{\Lambdac \to \proton \KS}\xspace}
\newcommand{\LctopKspipi}       {\ensuremath{\Lambdac \to \proton \KS \pip \pim}\xspace}
\newcommand{\LctoLambdapi}      {\ensuremath{\Lambdac \to \Lambda \pip}\xspace}
\newcommand{\LctoLambdapipipi}  {\ensuremath{\Lambdac \to \Lambda \pip \pim \pip}\xspace}

\newcommand{\mLcp}              {\ensuremath{m_{\Lambda_c \antiproton}}\xspace}
\newcommand{\mLcpi}             {\ensuremath{m_{\Lambda_c \pi}}\xspace}
\newcommand{\mppi}              {\ensuremath{m_{\antiproton \pi}}\xspace}

\newcommand{\msqLcpi}           {\ensuremath{m^{2}_{\Lambda_c \pi}}\xspace}
\newcommand{\msqppi}            {\ensuremath{m^{2}_{\antiproton \pi}}\xspace}

\newcommand{\DstartoDpi}        {\ensuremath{\Dstarp \to \Dz \pip}\xspace}
\newcommand{\DtoKpi}            {\ensuremath{\Dz \to \Km \pip}\xspace}

\newcommand{\BABARPubYear}    {08}
\newcommand{\BABARPubNumber}  {016}

\newcommand{\SLACPubNumber} {13341}
\newcommand{\LANLNumber} {0807.4974}

\def\figurebox#1#2#3{%
    \def\arg{#3}%
    \ifx\arg\empty
    {\hfill\vbox{\hsize#2\hrule\hbox to #2{\vrule\hfill\vbox to #1{\hsize#2\vfill}\vrule}\hrule}\hfill}%
    \else
    {\hfill\epsfbox{#3}\hfill}%
    \fi}

\begin{document}

\preprint{\babar-PUB-\BABARPubYear/\BABARPubNumber} 
\preprint{SLAC-PUB-\SLACPubNumber} 

\begin{flushleft}
\babar-PUB-\BABARPubYear/\BABARPubNumber\\
SLAC-PUB-\SLACPubNumber\\
arXiv:\LANLNumber\\[10mm]
\end{flushleft}

\title{
{\large \bf \boldmath
Measurements of $\BR(\BzbtoLcpb)$ and $\BR(\BmtoLcpbpi)$ and Studies of $\Lambdac\pim$ Resonances}
}

%
\author{B.~Aubert}
\author{M.~Bona}
\author{Y.~Karyotakis}
\author{J.~P.~Lees}
\author{V.~Poireau}
\author{E.~Prencipe}
\author{X.~Prudent}
\author{V.~Tisserand}
\affiliation{Laboratoire de Physique des Particules, IN2P3/CNRS et Universit\'e de Savoie, F-74941 Annecy-Le-Vieux, France }
\author{J.~Garra~Tico}
\author{E.~Grauges}
\affiliation{Universitat de Barcelona, Facultat de Fisica, Departament ECM, E-08028 Barcelona, Spain }
% ...
\collaboration{The \babar\ Collaboration}
\noaffiliation


\begin{abstract}
We present an investigation of the decays \BzbtoLcpb and \BmtoLcpbpi based on $383\times10^6$ $\FourS \to \BB$ decays recorded with the \babar\ detector. We measure the branching fractions of these decays; their ratio is $\BR(\BmtoLcpbpi) / \BR(\BzbtoLcpb) = 15.4 \pm 1.8 \pm 0.3$. The \BmtoLcpbpi process exhibits an enhancement at the $\Lambdac \antiproton$ threshold and is a laboratory for searches for excited charm baryon states. We observe the resonant decays \BmtoSigcpb and \BmtoSigcstarstarpb but see no evidence for \BmtoSigcstarpb. This is the first observation of the decay \BmtoSigcstarstarpb; however, the mass of the observed excited \Sigcz state is $(2846\pm8\pm10)\mevcc$, which is somewhat inconsistent with previous measurements. Finally, we examine the angular distribution of the \BmtoSigcpb decays and measure the spin of the \Sigmacz baryon to be $1/2$, as predicted by the quark model. 
\end{abstract}

\pacs{13.25.Hw, 13.60.Rj, 14.20.Lq}%

\maketitle

\section{INTRODUCTION}

Baryonic decays of \B mesons, which contain a heavy bottom quark and a light up or down quark, provide a laboratory for a range of particle physics investigations: trends in decay rates and baryon production mechanisms; searches for exotic states such as pentaquarks and glueballs~\cite{ref:babarBtoppbarK,ref:babarBtoDppbar}; searches for excited baryon resonances; examination of the angular distributions of \B-meson decay products to determine baryon spins; and measurements of radiative baryonic \B decays that could be sensitive to new physics through flavor-changing neutral currents~\cite{ref:radBdecays_exp,ref:belleBtoLpbargamma}. The latter measurements rely on improving our theoretical understanding of baryonic \B decays in general~\cite{ref:radBdecays_th,ref:radBdecays_bary_th}.

The inclusive branching fraction for baryonic \B decays is $(6.8 \pm 0.6)\%$~\cite{ref:Argus1992}, and many exclusive baryonic \B decay modes have been observed~\cite{ref:pdg2006}. If we order the measured decays by $Q$-value:
\begin{equation}
Q = m_{\B} - \sum_f m_{f},
\end{equation}
where $m_{f}$ is the mass of each daughter in the final state of the \B decay, we find that for each type of baryonic \B decay, the branching fractions decrease as the $Q$-value increases. The smallest measured branching fraction is of the order $10^{-6}$, which also corresponds to our experimental sensitivity for measuring these branching fractions. Potentially interesting \B-meson decays such as $\B \to \proton \antiproton$, $\B \to \Lambda \Lbar$, and $\B \to \Lambdac \Lbar_c^-$ have not yet been seen.

Theoretical approaches to calculating baryonic \B decays include pole models~\cite{ref:Jarfipolemodel, ref:Deshpandepolemodel}, diquark models~\cite{ref:Balldiquarkmodel}, and QCD sum rules~\cite{ref:ReindersQCDsumrule, ref:CZQCDsumrule}. Recently, theoretical calculations have focused on pole models, where the \B decay proceeds through an intermediate \b-flavored baryon state, which then decays weakly into one of the final state baryons~\cite{ref:chengpolemodel2003, ref:chengpolemodel2002}. However, it is not clear that the pole model is reliable for baryon poles, and the predictions given in the literature vary significantly. Perhaps the most satisfying theoretical interpretation of baryonic \B decay rates is the qualitative one proposed by Hou and Soni in 2001~\cite{ref:housoni}, who argue that \B decays are favored if the baryon and antibaryon in the final-state configuration are close together in phase space. A consequence is that decay rates to two-body baryon-antibaryon final states are suppressed relative to rates of three-body final states containing the same baryon-antibaryon system plus an additional meson. In the three-body case, the baryon and antibaryon can be in the favored configuration---close together in phase space---rather than back-to-back as in the two-body case.

In this paper, we investigate the decays \BzbtoLcpb and \BmtoLcpbpi~\cite{ref:ChargeConjImplied}. We investigate baryon production in \B decays by comparing the two-body (\BzbtoLcpb) and three-body (\BmtoLcpbpi) decay rates directly. The dynamics of the baryon-antibaryon (\Lcp) system in the three-body decay provide insight into baryon production mechanisms. Additionally, the \BmtoLcpbpi system is a laboratory for studying excited baryon states and is used to measure the spin of the \Sigmacz. This is the first measurement of the spin of this state.


\section{\babar\ DETECTOR AND DATA SAMPLE}

The measurements presented in this paper are based on $383\times10^6$ $\FourS \to \BB$ decays recorded with the \babar\ detector~\cite{ref:babarnim} at the \pep2 \epem asymmetric-energy \BF\ at the Stanford Linear Accelerator Center. At the interaction point, 9-\gev electrons collide with 3.1-\gev positrons at the \FourS resonance with a center-of-mass energy of 10.58\gevcc. 

Charged particle trajectories are measured by a five-layer silicon vertex tracker (SVT) and a 40-layer drift chamber (DCH) immersed in a $1.5$-T axial magnetic field. Charged particle identification is provided by ionization energy (\dedx) measurements in the SVT and DCH along with Cherenkov radiation detection by an internally reflecting ring-imaging detector (DRC).

Exclusive \B-meson decays are simulated with the Monte Carlo (MC) event generator \texttt{EvtGen}~\cite{ref:EvtGen}. Background continuum MC samples ($\epem \to \qqbar$, where $\q = \u, \d, \s, \c$) are simulated using \texttt{Jetset7.4}~\cite{ref:Jetset} to model generic  hadronization processes. Background MC samples of $\epem \to \BpBm$ and $\BzBzb$ are based on simulations of many exclusive \B decays (also using \texttt{EvtGen}). The large samples of simulated events are generated and propagated through a detailed detector simulation using the \texttt{GEANT4} simulation package~\cite{ref:Geant4}.

\section{CANDIDATE SELECTION}

We select candidates that are kinematically consistent with \BzbtoLcpb and \BmtoLcpbpi. For the decay mode \BzbtoLcpb, we reconstruct \Lambdac candidates in the \pKpi, \pKs, \pKspipi, and \Lambdapi decay modes, requiring the invariant mass of each \Lambdac candidate to be within $10\mevcc$ of the world average value~\cite{ref:pdg2006}. For \BmtoLcpbpi, we also reconstruct \Lambdac candidates in the \Lambdapipipi decay mode, and require all of the \Lambdac candidates to have an invariant mass within $12\mevcc$ of the world average value.

The \proton, \kaon, and $\pi$ candidates must be well-reconstructed in the DCH and are identified with likelihood-based particle selectors using information from the SVT, DCH, and DRC.

The \KS candidates are reconstructed from two oppositely charged pion candidates that come from a common vertex; $\Lambda$ candidates are formed by combining a proton candidate with an oppositely charged pion candidate that comes from a common vertex. The invariant mass of each \KS and $\Lambda$ candidate must be within $10\mevcc$ of the world average value~\cite{ref:pdg2006} and the flight significance (defined as the flight distance from the \Lambdac vertex in the $x-y$ plane divided by the measurement uncertainty) must be greater than 2. The mass of each \KS and $\Lambda$ candidate is then constrained to the world average value~\cite{ref:pdg2006}.

A mass constraint is applied to all of the \Lambdac candidates, and all \Lambdac daughter tracks must come from a common vertex. The \Lambdac candidates are then combined with an antiproton to form a \BzbtoLcpb candidate, or with an antiproton and a pion to form a \BmtoLcpbpi candidate. The daughters of each \B candidate must come from a common vertex, and the candidate with the largest $\chisq$ probability in each event is selected.

Additional background suppression is provided by information about the topology of the events. A Fisher discriminant~\cite{ref:fisher} is constructed based on the absolute value of the cosine of the angle of the \B candidate momentum vector with respect to the beam axis in the \epem center-of-mass (CM) frame, the absolute value of the cosine of the angle between the \B candidate thrust axis~\cite{ref:thrust} and the thrust axis of the rest of the event in the \epem CM frame, and the moments $L_0$ and $L_2$. The quantity $L_j$ is defined as $\sum_i p_i \left| \cos\theta_i \right|^j$, where $\theta_i$ is the angle with respect to the \B candidate thrust axis of the $i$th charged particle or neutral cluster in the rest of the event and $p_i$ is its momentum. The optimal maximum value of the Fisher discriminant is chosen separately for each \Lambdac and \B decay mode.

Kinematic properties of \B-meson pair production at the \FourS provide further background discrimination. We define a pair of observables, $m_m$ and $m_r$, that are uncorrelated and exploit these constraints:
\begin{equation}
\begin{split}
m_m &= \sqrt{\left(q_{\epem}-\hat{q}_{\Lcp(\pim)}\right)^2}\quad\textrm{and}\\
m_r &= \sqrt{\left(q_{\Lcp(\pim)}\right)^2} - m_{\B}.
\end{split}
\end{equation}
The variable $m_m$ is based on the apparent recoil mass of the unreconstructed \B meson in the event, where $q_{\epem}$ is the four-momentum of the \epem system and $\hat{q}_{\Lcp(\pim)}$ is the four-momentum of the reconstructed \B candidate after applying a mass constraint. The variable $m_r$ is the difference between the unconstrained mass of the reconstructed \B candidate and $m_{\B}$, the world average value of the mass of the \B meson~\cite{ref:pdg2006}. Signal events peak at $m_{\B}$ in $m_m$ and $0$ in $m_r$. This set of variables was first used in~\cite{ref:babarmmiss} and is chosen as an uncorrelated alternative to $\DeltaE = E^{*}_{\B}-\frac{1}{2}\sqrt{s}$ and the energy-substituted mass $\mes = \sqrt{\frac{1}{4}s - p^{*2}_{\B}}$ (where $s = q^2_{\epem}$ and the asterisk denotes the \epem rest frame), which exhibit a $\sim30\%$ correlation for \BmtoLcpbpi. 

The event selection criteria are optimized based on studies of sideband data (in the region $0.10 < m_r < 0.20\gevcc$) and simulated signal MC samples. The data in a signal region  (approximately $\pm2\sigma$ wide in $m_m$ and $m_r$) were blinded until the selection criteria were determined and the signal extraction procedure was specified and validated. \B candidates that satisfy $m_m>5.121\gevcc$ and $\left|m_r\right|<0.10\gevcc$ are used in the maximum likelihood fit.

\section{BACKGROUNDS}

The primary source of background for \BzbtoLcpb candidates is continuum $\epem \to \qqbar$ events. Backgrounds due to decays such as \BmtoLcpbpi, $\Bzb \to \Lambdac \antiproton \piz$, and $\Bm \to \Sigcz \antiproton, \Sigcz \to \Lambdac \pim$ are rejected by the criterion $\left|m_r\right|<0.10\gevcc$.

Approximately equal amounts of continuum $\epem \to \qqbar$ and $\epem \to \BB$ events make up the background for \BmtoLcpbpi events. Again, the requirement $\left|m_r\right|<0.10\gevcc$ rejects most of the contributions from such decays as $\Bzb \to \Lambdac \antiproton \pip \pim$ and $\Bm \to \Lambdac \antiproton \pim \piz$. Approximately $1\%$ of the background in the fit region is due to these four-body events, but they do not peak in $m_m$ and $m_r$. A small peaking background is present from $\Bzb \to \Sigcp \antiproton, \Sigcp \to \Lambdac \piz$ events, especially when the $\piz$ has low momentum. Based on a branching fraction measurement of the isospin partner decay $\BR(\BmtoSigcpb)=(3.7\pm0.7\pm0.4\pm1.0)\times 10^{-5}$~\cite{ref:belle2004}, where the uncertainties are statistical, systematic, and the uncertainty due to $\BR(\LctopKpi)$, respectively, we expect $11.5\pm2.5$ peaking background events in the signal region for \BmtoLcpbpi, \LctopKpi. A correction is applied and a systematic uncertainty is assigned to compensate for these events.

\section{DETECTION EFFICIENCY}

The detection efficiencies for \BzbtoLcpb and \BmtoLcpbpi signal events are determined from signal MC samples with $175,000$ to over $1,600,000$ events in each sample, depending on the \Lambdac decay mode. To account for inaccuracies in the simulation of the detector, each MC event is assigned a weight based on each daughter particle's momentum and angle. These weights are determined from studies comparing large pure samples of protons, kaons, and pions in MC samples and data. Small corrections ($0.4-1.6\%$) are also applied to account for tracking inefficiencies due to the displaced \KS and $\Lambda$ vertices. These corrections depend on the \KS and $\Lambda$ daughter trajectories' transverse momentum and angle, and the distance between the beam spot and the displaced vertex.

The detection efficiency ($\varepsilon_l$) for \BzbtoLcpb signal events in each \Lambdac decay mode ($l$) is determined from the number of signal events extracted from an extended unbinned maximum likelihood fit to signal MC events. These events pass the same selection criteria as applied to data. The fit is performed in two dimensions, $m_m$ and $m_r$. The probability distribution function (PDF) for the background consists of a threshold function~\cite{ref:ARGUS} in $m_m$ multiplied by a first-order polynomial in $m_r$; this is the same as the background PDF used in the fit to the \BzbtoLcpb data. The signal PDF consists of a Gaussian in $m_m$ multiplied by a modified asymmetric Gaussian with a tail parameter in $m_r$. The detection efficiencies in each \Lambdac decay mode are summarized in Table~\ref{tb:Efficiencies}.

\begin{table}
\caption{Detection efficiency for \BzbtoLcpb signal events, determined from signal Monte Carlo samples and separated by \Lambdac decay mode. The numbers correspond to the efficiency for \BzbtoLcpb (\BmtoLcpbpi), \Lambdac \to $f_l$, where $f_l$ is a given final state. The efficiencies quoted for the \BmtoLcpbpi decays are averaged across phase space.}
\begin{center}
\begin{tabular}{l@{$\quad\quad$}c@{$\quad\quad$}c} \hline \hline
 & \multicolumn{2}{c}{Efficiency for $\Lambdac \to f_l$} \\
\multicolumn{1}{c}{$f_l$} & $\BzbtoLcpb$ & $\BmtoLcpbpi$ \\ \hline
\pKpi & $22.9\%$ & $15.4\%$ \\
\pKs & $21.6\%$ & $14.3\%$ \\
\pKspipi & \phantom{0}$9.6\%$ & \phantom{1}$5.6\%$ \\
\Lambdapi & $17.2\%$ & $11.6\%$ \\ 
\Lambdapipipi & -- & \phantom{1}$4.0\%$ \\ \hline
\end{tabular}
\end{center}
\label{tb:Efficiencies}
\end{table}

The detection efficiency for \BmtoLcpbpi signal events in each \Lambdac decay mode varies considerably across the Dalitz plane of the three-body decay. For reference, we quote the average efficiencies in Table~\ref{tb:Efficiencies}, but we apply a more sophisticated treatment to these events. We parameterize the physical Dalitz region using the variables $\cos\theta_h$ and the $\Lambdac\,\pim$ invariant mass,  $m_{\Lambda_c\pi}$. The helicity angle $\theta_h$ is defined as the angle between the \pim and the \antiproton in the \Bm rest frame. The quantity $\cos\theta_h$ can be expressed in terms of Lorentz-invariant products of four-vectors. We divide the kinematic region into reasonably sized bins that are uniform in $\cos\theta_h$ ($0.2$ units wide) and nonuniform in $m_{\Lambda_c\pi}$ ($60-200\mevcc$ wide). This choice of variables is more conducive to rectangular bins than the traditional set of Dalitz variables. The $m_{\Lambda_c\pi}$ bins are narrower near the kinematic limits where the efficiency changes more rapidly and are centered on expected resonances. For \BmtoLcpbpi, \LctopKpi near $\cos\theta_h = 0$, the efficiency varies from approximately $13\%$ at low \mLcpi, to $16\%$ in the central \mLcpi region, to $8\%$ at high \mLcpi. The efficiency is fairly uniform with respect to $\cos\theta_h$, except at $\cos\theta_h\sim1$ and low \mLcpi, where it drops to $7.4\%$. The other \Lambdac decay modes exhibit similar variations in efficiency.

\section{SIGNAL EXTRACTION}

To extract the number of signal events in data, a two-dimensional ($m_m$ vs.\ $m_r$) extended unbinned maximum likelihood fit is performed simultaneously across \Lambdac decay modes. \BzbtoLcpb candidates and \BmtoLcpbpi candidates are fit separately. 

The background PDF for each fit is a threshold function~\cite{ref:ARGUS} in $m_m$ multiplied by a first-order polynomial in $m_r$. The shape parameter ($\vec{s}_{bkg}$) of the threshold function is free but is common to all of the \Lambdac decay modes. The slope $a$ of the first-order polynomial is allowed to vary independently for each \Lambdac decay mode.

The signal PDF is a single Gaussian distribution in $m_m$ multiplied by a single Gaussian distribution in $m_r$ for \BzbtoLcpb and multiplied by a double Gaussian distribution in $m_r$ for \BmtoLcpbpi. A single Gaussian is sufficient to describe the signal PDF for \BzbtoLcpb because of the small number of expected signal events. All of the shape parameters of the signal PDF ($\vec{s}_{sig}$) are free but are shared among the \Lambdac decay modes. Separate signal ($N_{sig,l}$) and background ($N_{bkg,l}$) yields are extracted for each \Lambdac decay mode $l$. 

The total likelihood is the product of the likelihoods for each \Lambdac decay mode:
\begin{equation}
{\cal L}_{tot} = \prod_{l}{\cal L}_l \left(\vec{y}_l;N_{sig,l},N_{bkg,l},\vec{s}_{sig},\vec{s}_{bkg},a_l\right).
\end{equation}
The symbol $\vec{y}$ represents the variables used in the 2-D fit, $\{m_m, m_r\}$.

The full simultaneous fit is validated using independent samples of signal MC events to simulate signal events and toy MC samples (generated from the background MC sample distribution) to represent background events in the fit region. For both \BzbtoLcpb and \BmtoLcpbpi, we perform fits to 100 combined MC samples and find that the fit is robust and the results are unbiased.

\begin{figure*}[tbp]
\begin{center}
\epsfig{file=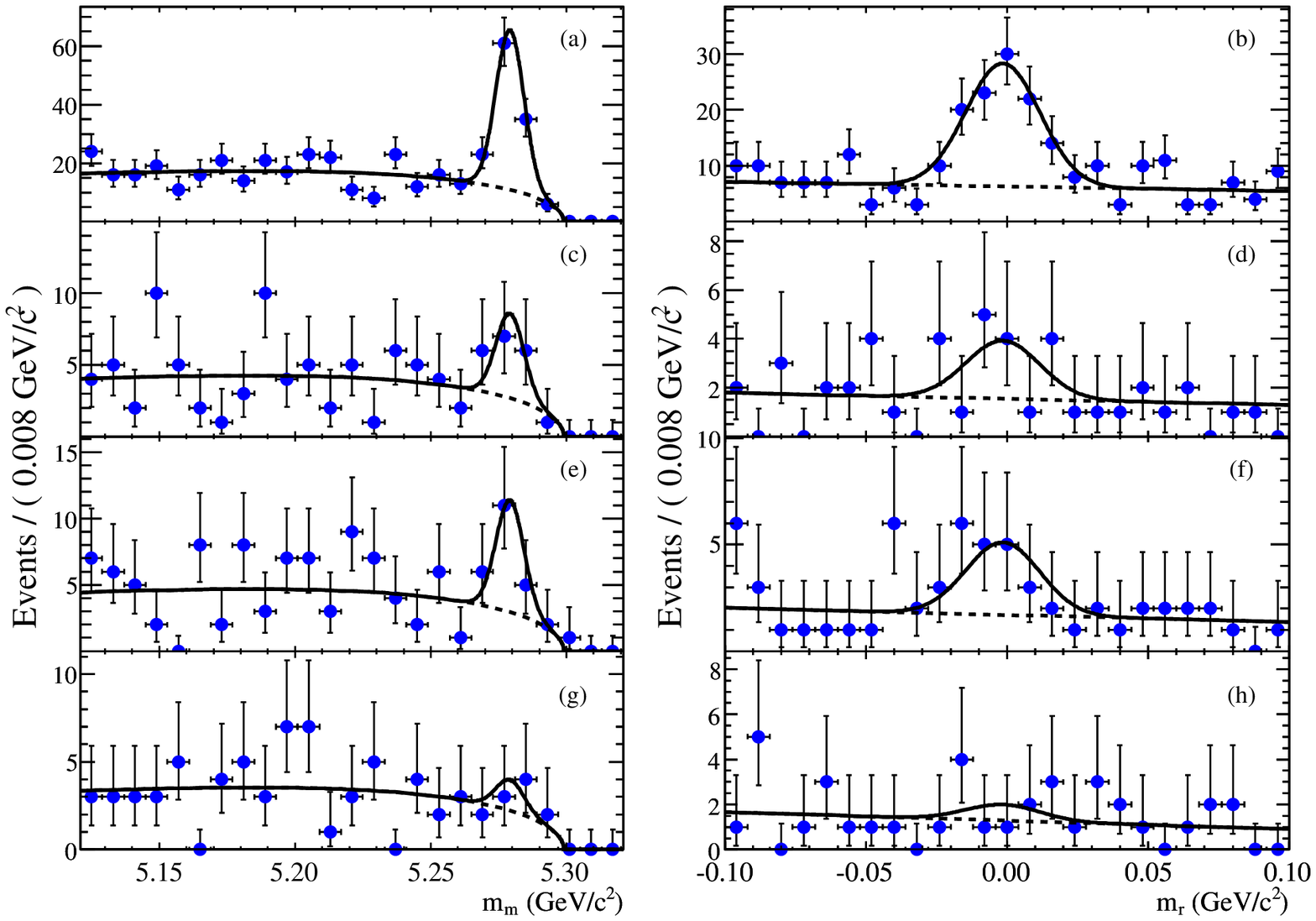,width=\textwidth}
\renewcommand{\baselinestretch}{1}
\caption{\label{fg:BzFitResults} Projections of $m_m$ (left) and $m_r$ (right) in data for \BzbtoLcpb candidates, separated by \Lambdac decay mode: $(a, b)$ are \LctopKpi, $(c, d)$ are \LctopKs, $(e, f)$ are \LctopKspipi, and $(g, h)$ are \LctoLambdapi. The $m_m$ projections $(a, c, e, g)$ are for $|m_r|<0.030\gevcc$ and the $m_r$ projections $(b, d, f, h)$ are for $m_m>5.27\gevcc$. The solid curves correspond to the PDF from the simultaneous 2-D fit to candidates for the four \Lambdac decay modes, and the dashed curves represent the background component of the PDF.}
\end{center}
\end{figure*}

\begin{figure*}[tbp]
\begin{center}
\epsfig{file=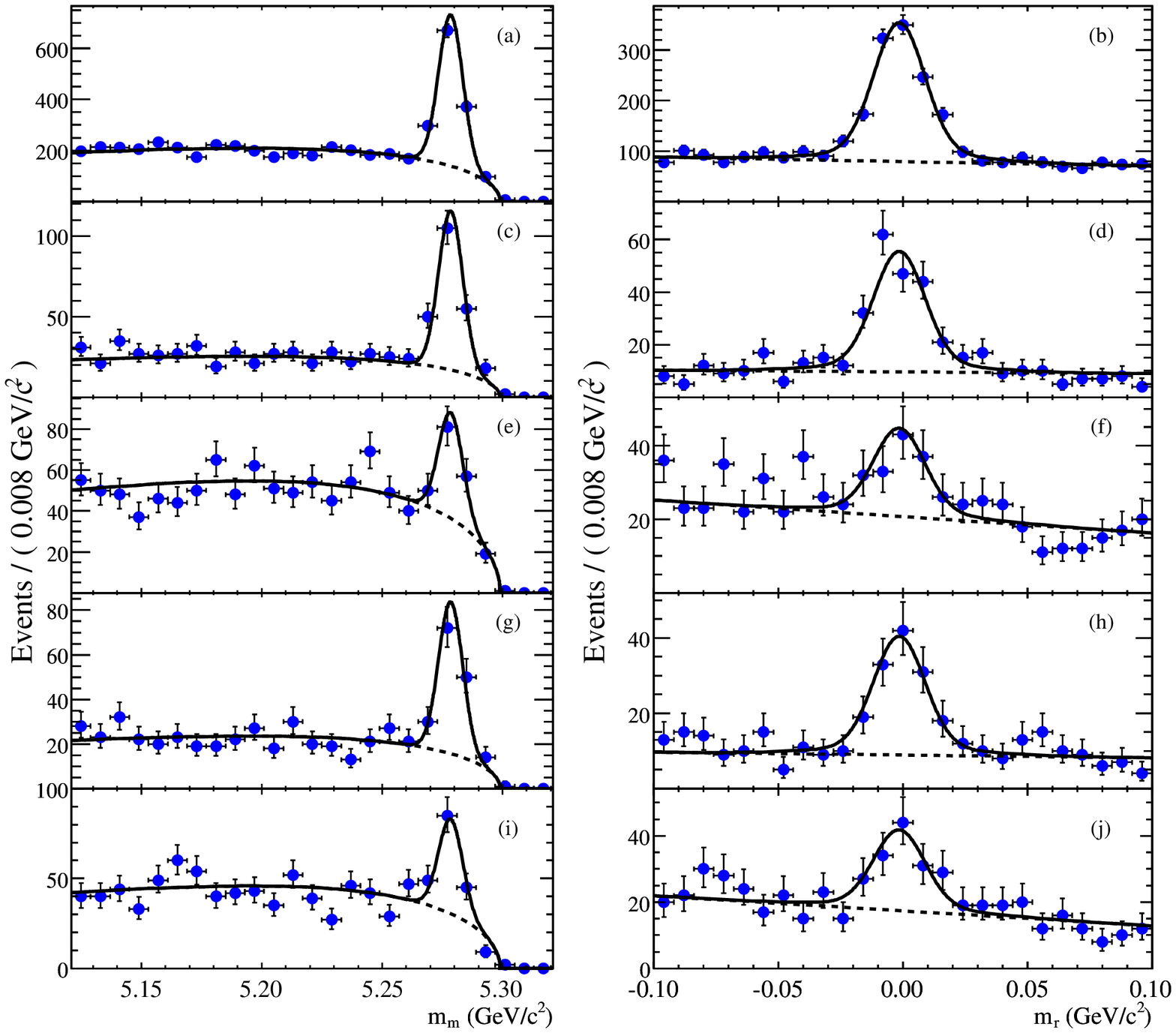,width=\textwidth}
\renewcommand{\baselinestretch}{1}
\caption{\label{fg:BmFitResults} Projections of $m_m$ (left) and $m_r$ (right) in data for \BmtoLcpbpi candidates, separated by \Lambdac decay mode: $(a, b)$ are \LctopKpi, $(c, d)$ are \LctopKs, $(e, f)$ are \LctopKspipi, $(g, h)$ are \LctoLambdapi, and $(i, j)$ are \LctoLambdapipipi. The $m_m$ projections $(a, c, e, g, i)$ are for $|m_r|<0.030\gevcc$ and the $m_r$ projections $(b, d, f, h, j)$ are for $m_m>5.27\gevcc$. The solid curves correspond to the PDF from the simultaneous 2-D fit to candidates for the five \Lambdac decay modes, and the dashed curves represent the background component of the PDF.}
\end{center}
\end{figure*}

The results of the 2-D fits to data are shown in projections of $m_m$ and $m_r$ for each \Lambdac decay mode. Figure~\ref{fg:BzFitResults} shows the result of the fit to \BzbtoLcpb candidates and Fig.~\ref{fg:BmFitResults} shows the result of the fit to \BmtoLcpbpi candidates. The signal yields from the fits are summarized in Table~\ref{tb:FitResults}.

\begin{table}
\caption{Signal yields from simultaneous fits (across \Lambdac decay modes) to \BzbtoLcpb and \BmtoLcpbpi candidates.}
\begin{center}
\begin{tabular}{l r@{ $\pm$ }l r@{ $\pm$ }l} \hline \hline
     & \multicolumn{4}{c}{$N_{sig}$} \\
Mode & \multicolumn{2}{c}{\BzbtoLcpb} & \multicolumn{2}{c}{\BmtoLcpbpi} \\ \hline
\pKpi & $  90$ & $ 11$ & $  991$ & $ 45$\\
\pKs & $  10$ & $ 4$ & $  165$ & $ 15$\\
\pKspipi & $  14$ & $ 5$ & $  86$ & $ 14$\\
\Lambdapi & $  3$ & $ 3$ & $  114$ & $ 13$\\
\Lambdapipipi & \multicolumn{2}{c}{--} & $  88$ & $ 13$\\ \hline
\end{tabular}
\end{center}
\label{tb:FitResults}
\end{table}

\section{BRANCHING FRACTION MEASUREMENTS}

For the three-body mode \BmtoLcpbpi, the efficiency variation across the Dalitz plane requires a correction for each signal event in order to extract the branching fraction for this mode. We use the \splot method~\cite{ref:splot} to calculate a weight for each event $e$ based on the 2-D fit to the variables $\vec{y}$. We have $N_s = 2$ species (signal and background) for each \Lambdac decay mode and define $f_{j,k}$ as the signal ($j,k=1$) or background ($j,k=2$) PDF. The \splot weights are calculated as
\begin{equation}
_s{\mathcal P}_n (\vec{y}_e) = \frac{\sum_{j=1}^{N_s}{\mathbf V}_{nj} f_j(\vec{y}_e)}{\sum_{k=1}^{N_s} N_k f_k(\vec{y}_e)},
\end{equation}
where $_{s}{\mathcal P}_{n}(\vec{y}_e)$ is the \splot weight for species $n$, ${\mathbf V}$ is the covariance matrix for signal and background yields, $f_{j,k}(\vec{y}_e)$ is the value of PDF $f_{j,k}$ for event $e$, and $\vec{y}_e$ is the $m_m$ and $m_r$ value for event $e$. The elements of the inverse of the covariance matrix ${\mathbf V}$ are calculated as follows:
\begin{equation}
{\mathbf V}^{-1}_{nj} = \frac{\partial^2 (-\ln {\mathcal L})}{\partial N_n \partial N_j}
                      = \sum^{N}_{e=1} \frac{f_n(\vec{y}_e)f_j(\vec{y}_e)}{\left(\sum_{k=1}^{N_s} N_k f_k(\vec{y}_e)\right)^2},
\end{equation}
where the sum is over the $N$ candidates. Note that in the calculation of the covariance matrix, the data is refit to the same simultaneous PDF described above, except that all fit parameters other than the yields are fixed to the values from the original fit.

We use these \splot weights to generate a signal or background distribution for any quantity that is not correlated with $m_m$ or $m_r$. The \splot formalism is easily extended to incorporate an efficiency correction for each candidate. Each candidate is assigned a weight of $1/\varepsilon$, where the efficiency $\varepsilon$ for an event is determined by its location in the $\cos\theta_h$ vs.\ $m_{\Lambda_c\pi}$ plane.

The branching fraction for \BzbtoLcpb for \Lambdac decay mode $l$ is calculated as follows:
\begin{multline}
\BR(\BzbtoLcpb)_{l} = \\
\frac{N_{sig,l}}{N_{\BB}\;\times\;\varepsilon_{l}\;\times\;\calR_{l}\;\times\;\BR(\LctopKpi)},
\end{multline}
where $N_{\BB}$ is the number of \BB events in the data sample and $\calR_l$ is the ratio of \Lambdac branching fraction for decay mode $l$ to $\BR(\LctopKpi)$, taking care to include the $\KS \to \pip \pim$ and $\Lambda \to \proton \pim$ branching fractions where applicable.

In order to determine the branching fraction for \BmtoLcpbpi, we take the product of the \splot weight and efficiency weight for each candidate and sum over all of the candidates in the fit region. We simplify the notation by using $_sW_i$ to denote the value of the signal \splot weight for event $i$ and include a $1\%$ correction for the peaking background due to $\Bzb \to \Sigcp \antiproton, \Sigcp \to \Lambdac \piz$:
\begin{multline}
\BR(\BmtoLcpbpi)_{l} = \\
\frac{0.99 \times \left(\sum\limits_i \dfrac{_sW_i}{\varepsilon_i}\right)_l}{N_{\BB}\;\times\;\calR_{l}\;\times\;\BR(\LctopKpi)}.
\end{multline}
The contribution from the peaking background is estimated using the \LctopKpi decay mode. Since the overall branching fraction for the peaking background contribution is the same regardless of \Lambdac decay mode, it is applied as a proportional correction. The measurements for $\BR(\BzbtoLcpb)$ and $\BR(\BmtoLcpbpi)$ for each \Lambdac decay mode are summarized in Table~\ref{tb:BFMeasurements}.

The BLUE (Best Linear Unbiased Estimate) technique is used as described in Ref.~\cite{ref:bluemethod} to combine the {\em correlated} branching fraction measurements for different \Lambdac decay modes. The purpose of the method is to obtain an estimate $\hat{x}$ that is a linear combination of $t$ individual measurements ($x_l$), is unbiased, and has the minimum possible variance $\hat{\sigma}^2$. The estimate $\hat{x}$ is defined by
\begin{equation}
\hat{x} = \sum_l \alpha_l x_l.
\label{eqn:xhat}
\end{equation}
The condition $\sum_l \alpha_l = 1$ ensures that the method is unbiased. Each coefficient $\alpha_l$ is a constant weight for measurement $x_l$ and is not necessarily positive. The set of coefficients $\boldsymbol{\alpha}$ (a vector with $t$ elements) is determined by
\begin{equation}
\boldsymbol{\alpha} = \frac{{\mathbf E}^{-1}\,{\mathbf U}}{{\mathbf U}_T\,{\mathbf E}^{-1}\,{\mathbf U}},
\end{equation}
where ${\mathbf U}$ is a $t$-component vector whose elements are all $1$ (${\mathbf U_T}$ is its transpose) and ${\mathbf E}$ is the ($t \times t$) error matrix. The diagonal elements of ${\mathbf E}$ are the individual variances, $\sigma^2_l$. The off-diagonal elements are the covariances between measurements ($r\sigma_l\sigma_{l^\prime}$, where $r$ is the correlation between measurements $l$ and $l^\prime$). The error matrices add linearly, so we define ${\mathbf E} = {\mathbf E}_{\textrm stat} + {\mathbf E}_{\textrm syst}$. ${\mathbf E}_{\textrm stat}$ includes the uncertainties in the fit yields and the correlations between yields from the simultaneous fit result. ${\mathbf E}_{\textrm syst}$ includes the systematic uncertainties that are described in the next section. Overall multiplicative constants ($N_{\BB}$ and $\BR(\LctopKpi)$) that are common to all the measurements and their uncertainties are not included in the BLUE method.

The solutions for $\boldsymbol{\alpha}$ are
\begin{multline}
\BzbtoLcpb:\\
\boldsymbol{\alpha}_T = \left(\begin{array}{cccc} 0.757 & 0.128 & 0.019 & 0.096\\ \end{array}\right),
\end{multline}
and
\begin{multline}
\BmtoLcpbpi:\\
\boldsymbol{\alpha}_T = \left(\begin{array}{ccccc} 0.913 & 0.043 & -0.003 & 0.029 & 0.018\\ \end{array}\right),
\end{multline}
where $\boldsymbol{\alpha}_T$ is the transpose of $\boldsymbol{\alpha}$. The order of the coefficients corresponds to the order of $\Lambdac$ decay modes presented in Table~\ref{tb:BFMeasurements}.

We calculate the best estimate $\hat{x}$ according to Eqn.~\ref{eqn:xhat} and divide this quantity by $N_{\BB}$ and $\BR(\LctopKpi)$. We calculate the variance of $\hat{x}$
\begin{equation}
\hat{\sigma}^2 = \boldsymbol{\alpha}_T\,{\mathbf E}\,\boldsymbol{\alpha}.
\end{equation}
Since the error matrices add linearly, we quote separate statistical and systematic uncertainties. The statistical and systematic uncertainties in $N_{\BB}$ are added in quadrature with the statistical and systematic $\hat{\sigma}$ results, respectively. The combined branching fraction measurements are thus
\begin{multline}
\BR(\BzbtoLcpb) =\\
\shoveright{(1.89 \pm 0.21 \pm 0.06 \pm 0.49) \times 10^{-5},}\\
\shoveleft{\BR(\BmtoLcpbpi) =}\\
 (3.38 \pm 0.12 \pm 0.12 \pm 0.88) \times 10^{-4},
\end{multline}
where the uncertainties are statistical, systematic, and the uncertainty in $\BR(\LctopKpi)$, respectively.

We use the same procedure to determine the branching ratio $\BR(\BmtoLcpbpi)/\BR(\BzbtoLcpb)$:
\begin{equation}
\frac{\BR(\BmtoLcpbpi)}{\BR(\BzbtoLcpb)} = 15.4 \pm 1.8 \pm 0.3.
\end{equation}
In the branching ratio, many of the systematic uncertainties, including the dominant $\BR(\LctopKpi)$ uncertainty, cancel.

\begin{table*}[tbp]
\caption{\label{tb:BFMeasurements} Summary of the individual and combined branching fraction measurements for \BzbtoLcpb and \BmtoLcpbpi. The uncertainties are statistical, systematic, and due to \BR(\LctopKpi), respectively.}
\begin{center}
\begin{tabular}{l c c c} \hline \hline
Mode & \BR(\BzbtoLcpb) & \BR(\BmtoLcpbpi) \\ \hline
\LctopKpi & $(2.05 \pm 0.25 \pm 0.05 \pm 0.53)\times10^{-5}$ & $(3.38 \pm 0.13 \pm 0.11 \pm 0.88)\times10^{-4}$\\
\LctopKs & $(1.49 \pm 0.60 \pm 0.17 \pm 0.39)\times10^{-5}$ & $(3.82 \pm 0.35 \pm 0.38 \pm 0.99)\times10^{-4}$\\
\LctopKspipi & $(4.33 \pm 1.55 \pm 0.57 \pm 1.13)\times10^{-5}$ & $(4.58 \pm 0.70 \pm 0.66 \pm 1.19)\times10^{-4}$\\
\LctoLambdapi & $(0.71 \pm 0.71 \pm 0.18 \pm 0.18)\times10^{-5}$ & $(3.98 \pm 0.45 \pm 0.39 \pm 1.03)\times10^{-4}$\\
\LctoLambdapipipi & -- & $(3.49 \pm 0.51 \pm 0.38 \pm 0.91)\times10^{-4}$\\
combined & $(1.89 \pm 0.21 \pm 0.06 \pm 0.49) \times 10^{-5}$ & $(3.38 \pm 0.12 \pm 0.12 \pm 0.88) \times 10^{-4}$\\ \hline
\end{tabular}
\end{center}
\end{table*}

\section{SYSTEMATIC UNCERTAINTIES IN THE BRANCHING FRACTIONS}

The uncertainties in the $\BR(\BzbtoLcpb)$ and $\BR(\BmtoLcpbpi)$ measurements are dominated by the uncertainty in the $\LctopKpi$ branching fraction, and then by the uncertainties in the \Lambdac branching ratios (compared to \LctopKpi)~\cite{ref:pdg2006}. 

The systematic uncertainties for each \Lambdac decay mode are summarized in Tables~\ref{tb:BzSystematics} and \ref{tb:BmSystematics}. The systematic uncertainty in the number of \BB\ events is $1.1\%$.

The uncertainty in the efficiency determination is due to MC statistics, charged particle tracking, and particle identification. For \BzbtoLcpb, the uncertainty due to MC statistics is $(0.4-0.6)\%$. For \BmtoLcpbpi, we calculate the uncertainty due to MC statistics by independently varying the number of reconstructed signal MC events in each $\cos\theta_h, m_{\Lambda_c\pi}$ bin according to a Poisson distribution (ensuring that the data events in the same bin are correlated). The resulting uncertainty is $(0.6-3.1)\%$.

The tracking efficiency systematic uncertainties are determined from two separate studies. In the first study, $\taup\taum$ candidates are selected, in which one $\tau$ candidate decays to leptons and the other decays to more than one hadron plus a neutrino. Events are selected if one lepton and at least two charged hadrons are reconstructed. The efficiency is then measured for reconstructing the third charged particle for the hadronic decay. From this study there is a $(0.38-0.45)\%$ uncertainty in the tracking efficiency per charged particle. In the second method, a charged particle trajectory is reconstructed in the SVT, and then the efficiency for finding the corresponding trajectory in the DCH is measured. For the latter study, the uncertainties range from $0.21\%$ to $1.18\%$ depending on the \Lambdac decay mode. The systematic uncertainties determined in the two studies are added in quadrature.

The systematic uncertainty for charged particle identification is a measure of how well the corrections applied to the events in the signal MC sample for the efficiency determination describe the \BzbtoLcpb and \BmtoLcpbpi decay modes. The corrections are determined from control MC and data samples (a $\Lambda \to \proton \pim$ sample for protons and $\DstartoDpi, \DtoKpi$ samples for pions and kaons). The efficiency as a function of momentum and angle for the \BzbtoLcpb and \BmtoLcpbpi signal MC samples and the control MC samples agree to within $(0.8-3.5)\%$ for different ranges of momentum and angle.

Systematic uncertainties due to the fit procedure are also considered. The dominant fit uncertainty is due to the threshold function parameter in the background PDF. The fit validation study showed that this parameter must be shared among \Lambdac decay modes to ensure fit robustness. We allowed this parameter to vary independently among \Lambdac decay modes and repeated the fit to data; the difference between the fit results is taken as a systematic uncertainty of $(1-24)\%$ depending on the \Lambdac decay mode. A peaking background due to misreconstructed $\Bzb \to \Sigcp \antiproton, \Sigcp \to \Lambdac \piz$ events is present at the level of $1.0\%$. We assign a systematic uncertainty of $0.5\%$ to account for the uncertainty in $\BR(\Bzb \to \Sigcp \antiproton)$. The nominal endpoint in the fit to $m_m$ is $5289.0\mevcc$; we vary this by $\pm0.5\mevcc$, resulting in a systematic uncertainty of $(0.2-1.5)\%$.

\begin{table*}[hbp]
\caption{Summary of the contributions to the relative systematic uncertainty in $\BR(\BzbtoLcpb)$ for each \Lambdac decay mode. The total for each mode is determined by adding the uncertainty from each source in quadrature. The fractional statistical uncertainty in the fit yield for each mode is provided for comparison.}
\begin{center}
\begin{tabular}{c c c c c} \hline \hline
\multirow{2}{*}{Source} & \multicolumn{4}{c}{\BzbtoLcpb\; Systematic Uncertainty} \\
 & \LctopKpi & \LctopKs & \LctopKspipi & \LctoLambdapi \\ \hline
\BB\ events & $\phantom{1}1.1\%$ & $\phantom{1}1.1\%$ & $\phantom{1}1.1\%$ & $\phantom{2}1.1\%$ \\
$\calR_l$ & -- & $\phantom{1}8.5\%$ & $11.8\%$ & $\phantom{2}8.9\%$ \\
MC statistics & $\phantom{1}0.4\%$ & $\phantom{1}0.6\%$ & $\phantom{1}0.6\%$ & $\phantom{2}0.4\%$  \\
Tracking & $\phantom{1}1.7\%$ & $\phantom{1}1.9\%$ & $\phantom{1}2.8\%$ & $\phantom{2}1.7\%$ \\
Displaced Vertices & -- & $\phantom{1}1.1\%$ & $\phantom{1}1.1\%$ & $\phantom{2}1.1\%$ \\
Particle Identification & $\phantom{1}1.5\%$ & $\phantom{1}2.1\%$ & $\phantom{1}1.7\%$ & $\phantom{2}1.6\%$ \\ 
Fitting & $\phantom{1}0.9\%$ & $\phantom{1}7.0\%$ & $\phantom{1}4.9\%$ & $24.2\%$ \\ \hline
Total & $\phantom{1}3\%$ & $\phantom{1}12\%$ & $\phantom{1}13\%$ & $\phantom{1}26\%$ \\ \hline
Statistical & $12\%$ & $\phantom{1}40\%$ & $\phantom{1}36\%$ & $100\%$ \\ \hline \hline
\end{tabular}
\end{center}
\label{tb:BzSystematics}
\end{table*}

\begin{table*}[tbp]
\caption{Summary of the contributions to the relative systematic uncertainty in $\BR(\BmtoLcpbpi)$ for each \Lambdac decay mode. The total for each mode is determined by adding the uncertainty from each source in quadrature. The fractional statistical uncertainty in the fit yield for each mode is provided for comparison.}
\begin{center}
\begin{tabular}{c c c c c c} \hline \hline
\multirow{2}{*}{Source} & \multicolumn{5}{c}{\BmtoLcpbpi\; Systematic Uncertainty} \\
 & \LctopKpi & \LctopKs & \LctopKspipi & \LctoLambdapi & \LctoLambdapipipi \\ \hline
\BB\ events & $1.1\%$ & $1.1\%$ & $\phantom{1}1.1\%$ & $\phantom{1}1.1\%$ & $\phantom{1}1.1\%$\\
$\calR_l$ & -- & $8.5\%$ & $11.8\%$ & $\phantom{1}8.9\%$ & $\phantom{1}6.1\%$\\
MC statistics & $0.6\%$ & $2.1\%$ & $\phantom{1}3.1\%$ & $\phantom{1}2.0\%$ & $\phantom{1}3.0\%$ \\
Tracking & $2.6\%$ & $2.3\%$ & $\phantom{1}3.2\%$ & $\phantom{1}2.1\%$ & $\phantom{1}2.9\%$ \\
Displaced Vertices & -- & $1.1\%$ & $\phantom{1}1.1\%$ & $\phantom{1}1.1\%$ & $\phantom{1}1.1\%$ \\
Particle Identification & $0.8\%$ & $1.8\%$ & $\phantom{1}2.5\%$ & $\phantom{1}1.8\%$ & $\phantom{1}3.5\%$ \\
Fitting & $1.6\%$ & $3.2\%$ & $\phantom{1}6.5\%$ & $\phantom{1}2.5\%$ & $\phantom{1}2.4\%$ \\ \hline
Total & $3.4\%$ & $9.9\%$ & $14.5\%$ & $10.0\%$ & $\phantom{1}8.7\%$ \\ \hline
Statistical & $4.5\%$ & $9.1\%$ & $16.3\%$ & $11.4\%$ & $14.8\%$ \\ \hline \hline
\end{tabular}
\end{center}
\label{tb:BmSystematics}
\end{table*}

\section{$\boldsymbol{\Lambdac \antiproton}$ THRESHOLD ENHANCEMENT}

\begin{figure*}[tbp]
\begin{center}
\epsfig{file=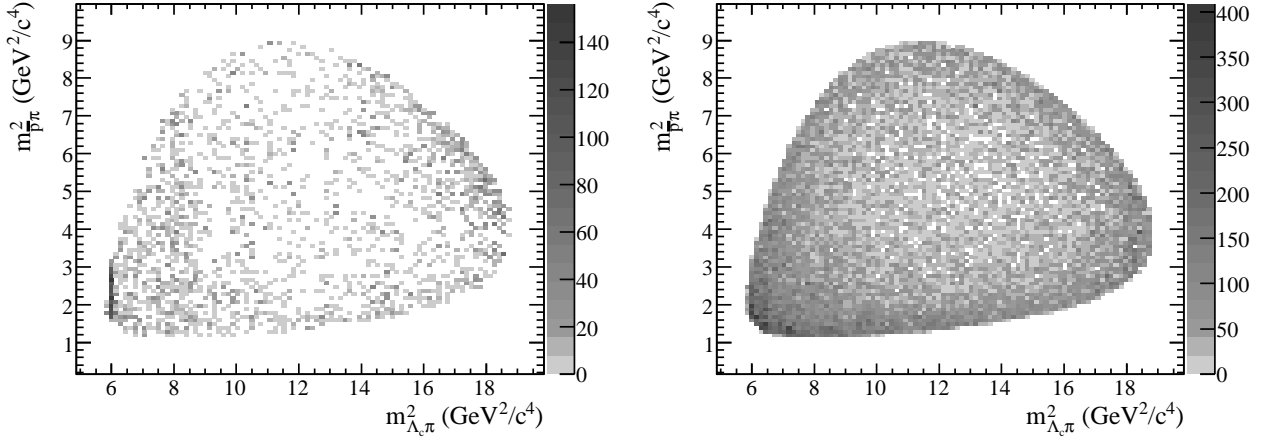,width=0.95\textwidth}
\renewcommand{\baselinestretch}{1}
\caption{\label{fg:Dalitzplot} Dalitz plot of $\msqppi$ vs. $\msqLcpi$. Each event is efficiency-corrected and given a signal (left) or background (right) \splot weight. Note that the density scales on the left- and right-hand plots are different, and negative bins are suppressed.}
\end{center}
\end{figure*}

The kinematic features and resonances in \BmtoLcpbpi are investigated through examination of the 2-D Dalitz plot ($\msqppi$ vs.\ $\msqLcpi$) and its projections (the resonances will be discussed in the next section). Using the \splot formalism, we project the events in the $\{m_m, m_r\}$ fit region using the signal \splot weights and background \splot weights along with the efficiency corrections. This method allows us to project only the features of the signal events, while taking the efficiency variations into account. Figure~\ref{fg:Dalitzplot} shows the \splot weights for $\msqppi$ vs.\ $\msqLcpi$. Note that the negative bins are suppressed in the 2-D Dalitz plot.

\begin{figure}[htbp]
\begin{center}
\epsfig{file=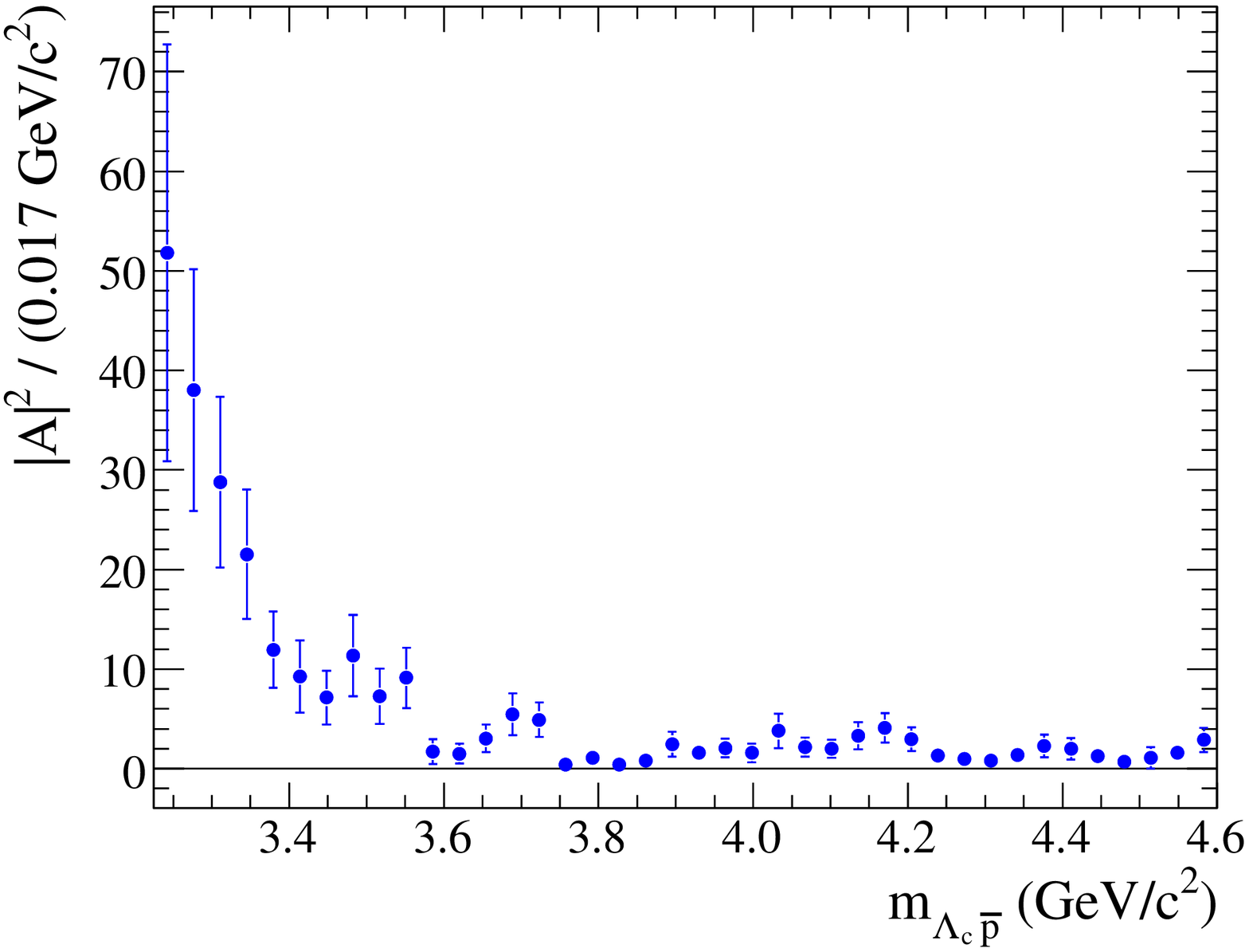,width=0.45\textwidth}
\renewcommand{\baselinestretch}{1}
\caption{\label{fg:LcpPS_CorrData} Projection of the amplitude squared ($|A|^2$) vs.\ \mLcp for \BmtoLcpbpi decays near threshold. Candidates are efficiency-corrected, weighted using the \splot technique, and corrected according to three-body phase space.}
\end{center}
\end{figure}

We project the events in the fit region onto the \mLcp axis with signal \splot weights and efficiency corrections to study the enhancement at threshold in the baryon-antibaryon mass distribution. This enhancement can be seen in \BmtoLcpbpi decays as a peak in \mLcp near the kinematic threshold, $m^0_{\Lambda_c\antiproton} = 3224.8\mevcc$. We divide the normalized \mLcp distribution by the expectation from three-body phase-space; the resulting distribution is shown in Figure~\ref{fg:LcpPS_CorrData}. An enhancement is clearly visible near threshold. The observation of this enhancement is consistent with baryon-antibaryon threshold enhancements as seen in other decay modes such as $\B \to \proton \antiproton \kaon$, $\B \to D \proton \antiproton (\pi)$, or in continuum $\epem \to \proton \antiproton \gamma$~\cite{ref:babarBtoppbarK,ref:babarBtoDppbar}.  

\section{RESONANT SUBSTRUCTURE OF $\boldsymbol{\BmtoLcpbpi}$}

We also project the events in the fit region onto the \mLcpi axis with signal \splot weights and efficiency corrections to study resonances in \mLcpi. We perform 1-D binned \chisq fits to discriminate between resonant (\Bm \to \Sigcz \antiproton) and nonresonant (\BmtoLcpbpi) signal events. In each binned \chisq fit, the PDF is numerically integrated over each (variable-sized) bin and the following quantity is minimized:
\begin{equation}
\chisq = \sum^{n_{\textrm{bins}}}_{i=1}\left(\frac{\int \left(Y_{\textrm{sig}} {\cal P}_{\textrm{sig}} + Y_{\textrm{nr}} {\cal P}_{\textrm{nr}}\right)\,dm_i\;-\;Y_i}{\sigma_i}\right)^2,
\end{equation}
where ${\cal P}_{\textrm{sig}}$ is the resonant signal PDF, ${\cal P}_{\textrm{nr}}$ is the nonresonant signal PDF, $Y_{\textrm{sig}}$ is the expected yield of weighted resonant signal events, and $Y_{\textrm{nr}}$ is the expected yield of weighted nonresonant signal events. We assume the amplitude and phase of the non-resonant \BmtoLcpbpi contribution is constant over the Dalitz plot, and does not interfere with the resonant contributions. The range of the integral over the quantity $dm_i$ takes into account the variable bin width, $Y_i$ is the number of weighted data events, and $\sigma_i$ is the uncertainty in $Y_i$. Variable bin widths are used to ensure that there are a sufficient number of signal events in each \mLcpi bin so that the estimated uncertainty is valid. This is especially important in the non-resonant sideband regions. Bin widths in the signal regions are chosen to have sufficient granularity throughout the resonant peaks.

The \Sigmacz and \Sigmaczstar are well-established resonances that decay to $\Lambdac \pim$. A third \Sigmac resonance, the \Sigmaczstarstar,  was reported by the Belle Collaboration in 2005~\cite{ref:BelleSigc2800} along with its isospin partners \Sigmacpstarstar and \Sigmacppstarstar. These resonances were observed in continuum ($\epem \to \ccbar$) $\Lambda_c \pi$ events. The \Sigmaczstarstar resonance was fit with a $D$-wave Breit-Wigner distribution and the mass difference $\Delta m = m_{\Sigmaczstarstar} - m_{\Lambdac} = (515~\pm~3^{~+~2}_{~-~6})\mevcc$ was measured, which corresponds to an absolute mass of $(2802^{~+~4}_{~-~7})\mevcc$~\cite{ref:pdg2006}. The natural width of the resonance is $(61^{+~28}_{-~18})\mev$~\cite{ref:BelleSigc2800}. We search for evidence of all three resonances.

\begin{figure}[htbp]
\begin{center}
\subfigure{\epsfig{file=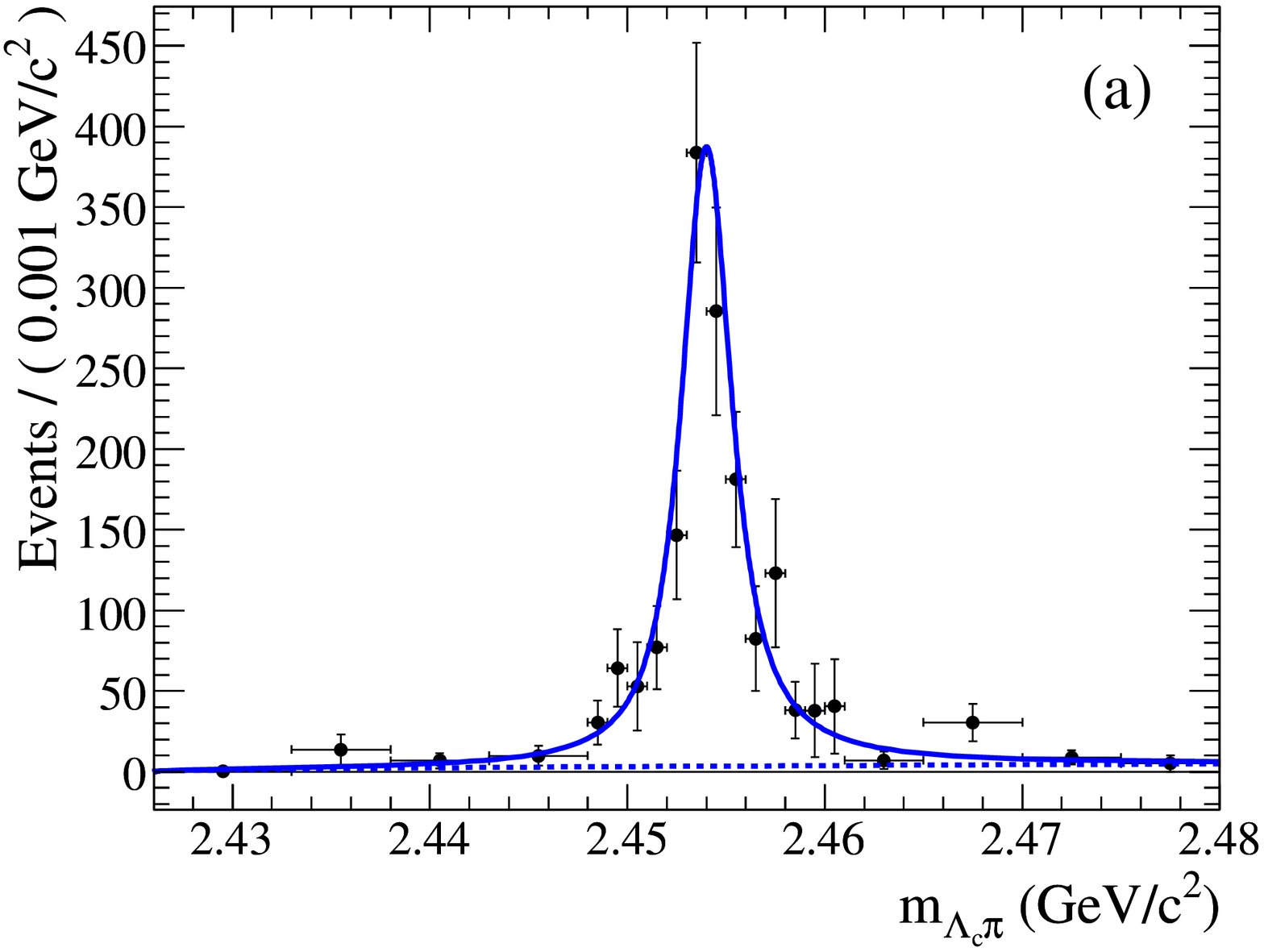,width=0.45\textwidth}}
\subfigure{\epsfig{file=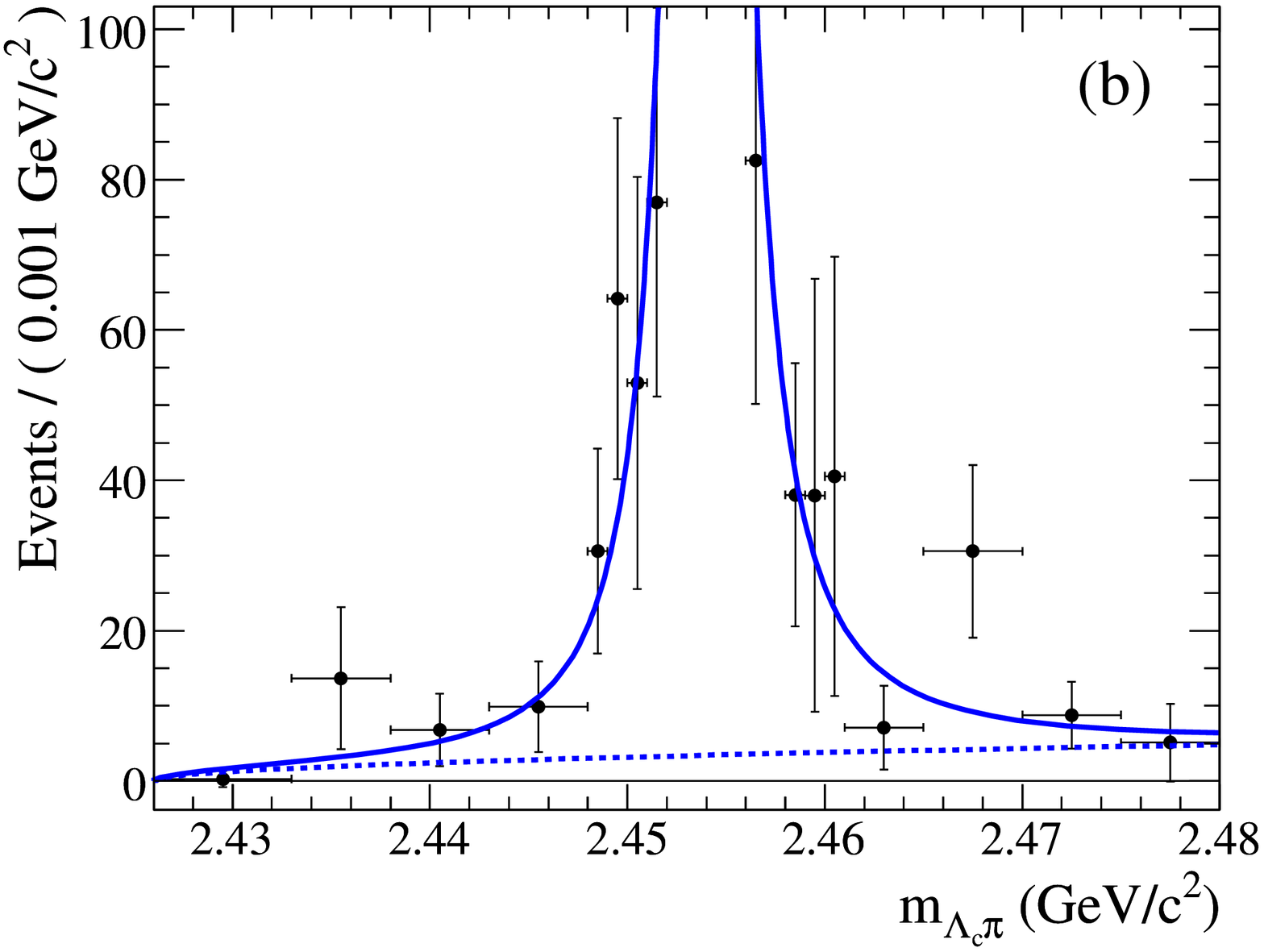,width=0.45\textwidth}}
\renewcommand{\baselinestretch}{1}
\caption{\label{fg:Sigc2455} (a) Projection of \mLcpi showing the \Sigmacz resonance. Events are efficiency corrected and weighted using the \splot technique, and the result of a binned \chisq fit to a Voigtian signal plus a threshold function background is overlaid. The variable bin sizes range from $1$ to $7\mevcc$. (b) The same fit result and data is shown on a smaller vertical scale to show the behavior of the PDF at threshold.}
\end{center}
\end{figure}

A significant \Sigmacz signal is seen near threshold; see Fig.~\ref{fg:Dalitzplot} and Fig.~\ref{fg:Sigc2455}. We construct a resonant signal PDF from a non-relativistic $S$-wave Breit-Wigner distribution convolved with the sum of two Gaussian distributions (to form a ``Voigtian" distribution). This quantity is multiplied by a phase-space function. The mass and (constant) width of the resonance in the Breit-Wigner PDF are free in the fit. The resolution (described by the two Gaussians) is fixed; it is determined from a $\BmtoSigcpb$, $\SigctoLcpi$, $\LctopKpi$ signal MC sample by comparing the measured \Sigmacz mass to the true \Sigmacz mass for each candidate. The two-body phase-space function goes to zero at the kinematic threshold for $\Lambda_c \pi$: $2426.03\mevcc$. The non-resonant signal PDF is a threshold function~\cite{ref:ARGUS}; the threshold is set to the kinematic threshold for $\Lambda_c \pi$. The shape parameter for the threshold function is fixed from a fit to a non-resonant \BmtoLcpbpi signal MC sample. The weighted data points are shown in Fig.~\ref{fg:Sigc2455} with the fit overlaid; the fit results are summarized in Table~\ref{tb:Sigc2455}. The average efficiency for \LctopKpi signal MC events in this region is $14.1\%$.

\begin{table}[htbp]
\caption{Fit results for \BmtoSigcpb. $Y_{\textrm{sig}}$ is the efficiency-corrected resonant signal yield in the fit range. Systematic uncertainties from the fit to $m_m$ vs.\ $m_r$ are not included in the yield. The world average values from the Particle Data Group (PDG) of the mass and width of the \Sigmacz are included for comparison~\cite{ref:pdg2006}.}
\begin{center}
\begin{tabular}{l r@{ $\pm$ }l c} \hline \hline
Fit Parameter & \multicolumn{2}{c}{Value} & PDG Value~\cite{ref:pdg2006} \\ \hline
$Y_{\textrm{sig}}$ & $1522$ & $149$ & \\
$m_R $ (\gevcc) & $2.4540$ & $ 0.0002$ & $2.4538 \pm 0.0002$\\
$\Gamma_R$ (\mev) & $2.6$ & $0.5$ & $2.2 \pm 0.4$ \\ \hline \hline
\end{tabular}
\end{center}
\label{tb:Sigc2455}
\end{table}

\begin{figure}[htbp]
\begin{center}
\epsfig{file=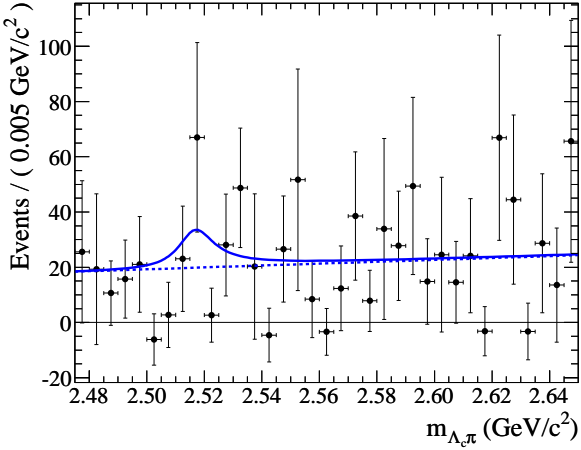,width=0.45\textwidth}
\renewcommand{\baselinestretch}{1}
\caption{\label{fg:Sigc2520} Projection of \mLcpi in the region of the \Sigmaczstar resonance. Events are efficiency corrected and weighted using the \splot technique, and the result of a binned \chisq fit to a relativistic $D$-wave Breit-Wigner signal with a mass-dependent width plus a linear background is overlaid. The bin size is $5\mevcc$. No significant signal is seen.}
\end{center}
\end{figure}

No significant signal is seen in the region of the \Sigmaczstar; see Fig.~\ref{fg:Sigc2520}. We perform a fit using a relativistic $D$-wave Breit-Wigner distribution with a mass-dependent width to describe the resonant signal PDF, fixing the resonance mass and the width to the world average values~\cite{ref:pdg2006}: $m_R = (2518.0 \pm 0.5)\mevcc$ and $\Gamma_R = (16.1 \pm 2.1)\mev$. The non-resonant signal PDF is a first-order polynomial. We obtain $Y_{\textrm{sig}} = 27 \pm 69$ events; the fit result is shown in Fig.~\ref{fg:Sigc2520}. The average efficiency for \LctopKpi signal MC events in this region is $15.4\%$.

In the \mLcpi distribution, we also observe an excited \Sigcz state that is higher in mass than the \Sigmaczstar. We construct a resonant signal PDF from a relativistic Breit-Wigner distribution with a mass-dependent width:
\begin{equation}
\Gamma(q) = \Gamma_R \left(\frac{q}{q_R}\right)^{2L+1} \left(\frac{m_R}{\mLcpi}\right) B^{\prime}_L(q,q_R)^2.
\label{eqn:Gamma_massdep}
\end{equation}
The quantity $L$ is the angular momentum ($L=0,1,2$ is $S$-wave, $P$-wave, $D$-wave, respectively), $q$ is the momentum of the \Lambdac (which is equal to the momentum of the \pim) in the excited \Sigcz rest frame, and $q_R$ is the value of $q$ when $\mLcpi=m_R$. In Eqn.~\ref{eqn:Gamma_massdep}, $B^{\prime}_L(q,q_R)$ is the Blatt-Weisskopf barrier factor~\cite{ref:pdg2006}: 
\begin{equation}
\begin{split}
B^{\prime}_0(q,q_R) &= 1,\\
B^{\prime}_1(q,q_R) &= \sqrt{\frac{1+q_R^2\,d^2}{1+q^2\,d^2}},\\
B^{\prime}_2(q,q_R) &= \sqrt{\frac{\left(q_R^2\,d^2 - 3\right)^2+9q_R^2\,d^2}{\left(q^2\,d^2 - 3\right)^2+9q^2\,d^2}},\\
\end{split}
\end{equation}
where we define a constant impact parameter $d = 1\fm$ (the approximate radius of a baryon), which corresponds to $5.1\gev^{-1}$. Blatt-Weisskopf barrier factors are weights that account for the fact that the maximum angular momentum ($L$) in a strong decay is limited by the linear momentum ($q$). Since the resonance is quite wide, we do not need to include a resolution function in the resonant signal PDF. The two fit parameters ($m_R$ and $\Gamma_R$) of the \Sigcz are free in the fit. The non-resonant signal PDF is a first-order polynomial.

\begin{figure}[tbp]
\begin{center}
\epsfig{file=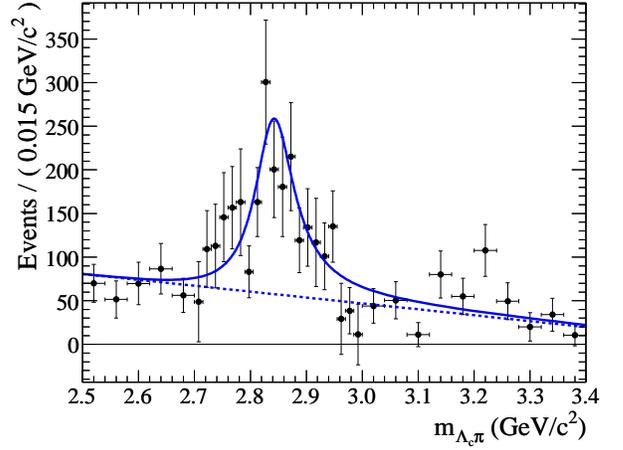,width=0.45\textwidth}
\renewcommand{\baselinestretch}{1}
\caption{\label{fg:ExcitedSigcz} Projection of \mLcpi showing an excited \Sigcz resonance. Events are efficiency corrected and weighted using the \splot technique. The result of a binned \chisq fit to a relativistic $D$-wave Breit-Wigner signal with a mass-dependent width plus a linear background is overlaid. The variable bin sizes range from $15$ to $40\mevcc$.}
\end{center}
\end{figure}

From the 1-D binned \chisq fit, we obtain $Y_{\textrm{sig}} = 1449 \pm 284$ efficiency-corrected events for the excited \Sigcz state. We choose $L=2$ for the nominal fit, but investigate $L=0,1$ as well. The average efficiency for \LctopKpi signal MC events in this region is $16.3\%$. The fit results are summarized in Table~\ref{tb:ExcitedSigcz} and shown in Fig.~\ref{fg:ExcitedSigcz}. The $\chisq$ from the fit is $37$ with $31$ degrees of freedom (DOF). If the signal yield is fixed to zero and the mean and width are fixed to the central values from the nominal fit, the resulting $\chisq$ is $78$ with $34$ DOF. The significance can be calculated from $\Delta \chisq = 40.9$, which is equivalent to $5.8\sigma$ for the joint estimation of three parameters.

The measured width of this state $(86^{~+~33}_{~-~22}\pm12)\mev$ is consistent with the width of the \Sigmaczstarstar measured by Belle~\cite{ref:BelleSigc2800}. However, the measured mass of this excited \Sigcz is $(2846 \pm 8 \pm 10)\mevcc$, which is $40\mevcc$ and $3\sigma$ higher (assuming Gaussian statistics) than Belle's measured mass for the \Sigmaczstarstar.

\begin{table}[tbp]
\caption{Fit results for the excited \Sigcz resonance. $Y_{\textrm{sig}}$ is the efficiency-corrected resonant signal yield in the fit range. Systematic uncertainties from the fit to $m_m$ vs.\ $m_r$ are not included in the yield. The world average values from the Particle Data Group (PDG) of the mass and width of the \Sigmaczstarstar are included for comparison~\cite{ref:pdg2006}.}
\begin{center}
\begin{tabular}{l r@{ $\pm$ }l c} \hline \hline
Fit Parameter & \multicolumn{2}{c}{Value} & PDG Value~\cite{ref:pdg2006} \\ \hline
$Y_{\textrm{sig}}$ & $1449$ & $284$ & \\
$m_R$ (\gevcc) & $2.846$ & $ 0.008$ & $2.802^{~+~0.004}_{~-~0.007}$ \\
$\Gamma_R$ (\mev) & \multicolumn{2}{c}{$86^{~+~33}_{~-~22}$} & $61^{~+~28}_{~-~18}$ \\ \hline \hline
\end{tabular}
\end{center}
\label{tb:ExcitedSigcz}
\end{table}

We evaluate systematic uncertainties for the \Sigmacz and the excited \Sigcz yields, masses, and widths by modifying the binning, the resonant signal PDF shape, and the non-resonant signal PDF shape. Changing the variable bin sizes leads to the dominant systematic uncertainty in the masses, widths, and yields of both resonances. For the \Sigmacz, the bin width in the peak region was decreased from the nominal $1\mevcc$ to $0.5\mevcc$; for the excited \Sigcz, the bin width was varied from $10$ to $20\mevcc$ compared to the nominal $15\mevcc$. For both resonances, an $S$-wave and a $P$-wave relativistic Breit-Wigner (without a resolution function) was used instead of the nominal resonant signal PDF. The (fixed) non-resonant threshold parameter for the \Sigmacz was varied by $\pm1\sigma$. A second-order polynomial was used (instead of a first-order polynomial) for the excited \Sigcz non-resonant PDF. A summary of the systematic uncertainties for $Y_{\textrm{sig}}$, $m_R$, and $\Gamma_R$ are summarized in Table~\ref{tb:ResSystematics}. 

\begin{table}
\caption{Systematic uncertainties for $Y_{\textrm{sig}}$, $m_R$ (in \mevcc), and $\Gamma_R$ (in \mev) for the \Sigmacz and excited \Sigcz resonances.}
\begin{center}
\begin{tabular}{c c c c c c} \hline \hline
 & \multicolumn{2}{c}{\Sigmacz} & \multicolumn{3}{c}{excited \Sigcz} \\
Systematic Source & $Y_{\textrm{sig}}$ & $\Gamma_R$ & $Y_{\textrm{sig}}$ & $m_R$ & $\Gamma_R$ \\ \hline
Resonant Signal PDF & -- & -- & $5.9\%$ & -- & $\pm\phantom{1}7$ \\
Non-resonant Signal PDF & -- & -- & $1.2\%$ & -- & $\pm\phantom{1}2$ \\
Binning & $6.9\%$ & $\pm0.3$ & $20\%$ & $\pm10$ & $\pm10$ \\ \hline
Total & $6.9\%$ & $\pm0.3$ & $21\%$ & $\pm10$ & $\pm12$ \\ \hline \hline
\end{tabular}
\end{center}
\label{tb:ResSystematics}
\end{table}

The significance is recalculated following each of the variations used to evaluate the systematic uncertainties in the excited \Sigcz resonance parameters. The resulting significance (including systematics) is $5.2\sigma$. A cross-check is performed to make sure the \Sigmaczstarstar signal is not the result of interference with a $\Delta(1232)^{++}$, for example (although no significant $\Delta(1232)^{++}$ signal is seen in the $\mppi$ distribution). The fit is performed again in the \Sigmaczstarstar mass region for candidates with $\mppi>1.5\gevcc$. We obtain $1329 \pm 230$ resonant signal events (compared to $1449 \pm 284$ events for the nominal fit) and a consistent mass and width.

An additional cross-check is performed to investigate whether there are appropriate fractions of resonant \Sigmaczstarstar events in different \Lambdac decay modes. This is accomplished by dividing the \splot-weighted, efficiency-corrected data into two samples according to the \Lambdac decay mode. Note that this cross-check neglects statistical correlations from the combined $m_r$ vs.\ $m_m$ fit (less than $15\%$) among the \Lambdac decay modes. A binned \chisq fit to only \LctopKpi candidates gives $Y_{\textrm{sig}} = 776 \pm 160$, compared to $6463 \pm 241$ total non-resonant \BmtoLcpbpi, \LctopKpi events ($(12 \pm 3)\%$). A binned \chisq fit to a combined sample of \LctopKs, \LctopKspipi, \LctoLambdapi, and \LctoLambdapipipi candidates gives $Y_{\textrm{sig}} = 530 \pm 177$ compared to $5956 \pm 431$ non-resonant events ($(9 \pm 3)\%$). (In order for this fit to converge, $m_R$ and $\Gamma_R$ were fixed to their nominal values.) The fractions are consistent in the two samples and the total ($1306 \pm 239$ events) is consistent with the nominal fit result within uncertainties.

We have also investigated the possibility that there are two resonances in the mass range shown in Figure~\ref{fg:ExcitedSigcz}. However, there is no evidence for two distinct vertical bands in this region in the \BmtoLcpbpi Dalitz plot (Figure~\ref{fg:Dalitzplot}), and we do not obtain a statistically significant fit to two resonances.

In order to measure the fraction of \BmtoLcpbpi decays that proceed through intermediate \Sigmac resonances, we assume that the contribution from each \Lambdac decay mode for events in the \Sigmac regions is equal to the measured contribution from each \Lambdac decay mode in all (resonant and non-resonant) \BmtoLcpbpi events. We set a $90\%$ C.L.\ upper limit on \BmtoSigcstarpb that includes systematic uncertainties and corresponds to $109$ events. The measured fractions or upper limits of \BmtoLcpbpi decays that proceed through an intermediate \Sigmac resonance are
\begin{equation}
\begin{split}
\frac{\BR(\BmtoSigcpb)}{\BR(\BmtoLcpbpi)}         &= (12.3 \pm 1.2 \pm 0.8) \times 10^{-2}, \\
\frac{\BR(\BmtoSigcstarstarpb)}{\BR(\BmtoLcpbpi)} &= (11.7 \pm 2.3 \pm 2.4) \times 10^{-2}, \\
\frac{\BR(\BmtoSigcstarpb)}{\BR(\BmtoLcpbpi)} &< 0.9 \times 10^{-2}\, (90\%\textrm{ C.L.}).
\end{split}
\end{equation}
Therefore approximately $1/4$ of \BmtoLcpbpi decays proceed through a known intermediate \Sigmac resonance.

\section{MEASUREMENT OF THE $\boldsymbol{\Sigmacz}$ SPIN}

The \Sigmacz is the lowest mass \Sigmac state. In the quark model, it is expected to have $J^P = \frac{1}{2}^+$, where $J$ is the spin and $P$ is the parity. In this section, we provide a quantitative evaluation of the spin-$1/2$ and spin-$3/2$ hypotheses for the \Sigmacz baryon.

We determine the spin of the \Sigmacz through an angular analysis of the decay \BmtoSigcpb, \SigctoLcpi. We define a helicity angle $\theta_h$ as the angle between the momentum vector of the \Lambdac and the momentum vector of the recoiling \B-daughter \antiproton in the rest frame of the \Sigmacz. If we assume $J(\Lambdac)=1/2$, the angular distributions for the spin-$1/2$ and spin-$3/2$ hypotheses for the \Sigmacz are
\begin{equation}
\begin{split}
J(\Sigcz)=\frac{1}{2}&:\quad \frac{dN}{d\cos\theta_h} \propto 1\\
J(\Sigcz)=\frac{3}{2}&:\quad \frac{dN}{d\cos\theta_h} \propto 1 + 3\cos^2\theta_h.
\end{split}
\label{eqn:AngDists}
\end{equation}
These are the ideal distributions; the measured angular distributions will be somewhat degraded due to nonuniform detector efficiencies, finite experimental resolution for measuring $\theta_h$, and background contamination. We estimate the effects of inefficiencies and background contamination by performing parameterized MC studies to quantify the decrease in sensitivity to discriminate between possible spin values. 

If we select from a $\pm2\sigma$ signal region in $m_m$ and $m_r$, without \splot weights or efficiency corrections, there are $127$ events in the \Sigmacz signal region and $27$ events in the \Sigmacz background regions ($2.430<\mLcpi < 2.445 \gevcc$ and $2.463< \mLcpi < 2.478 \gevcc$). Scaling the number of events in the background region by the ratio of the total width of the background regions compared to the width of the signal region, we expect $7.2\pm1.3$ background events in the signal region.

We measure the finite experimental resolution of $\cos\theta_h$ by comparing the measured value of $\cos\theta_h$ to the true value of $\cos\theta_h$ in $\BmtoLcpbpi, \LctopKpi$ events in a signal MC sample. The maximum root mean square of the measured value of $\cos\theta_h$ minus the true value of $\cos\theta_h$ in the \Sigmacz signal region determines the helicity angle resolution $\sigma(\cos\theta_h)<0.03$. Therefore the finite experimental resolution is small compared to any features in the spin-$1/2$ or spin-$3/2$ angular distributions.

We investigate the discrimination power between spin hypotheses using parameterized MC studies. In general, the log likelihood is computed as $\ln\calL = \sum_i w_i \ln(y_i)$, where $y_i$ is the probability density for observing event $i$. The weight $w_i$ for the MC studies is $w_i = \varepsilon_i$, where $\varepsilon_i$ is the efficiency for event $i$. We compute a log likelihood for each hypothesis:
\begin{equation}
\begin{split}
\ln \calL(1/2) =& \sum_i w_i \ln\frac{1}{2} \\
\ln \calL(3/2) =& \sum_i w_i \ln\left[\frac{1}{4}\left(1+3\cos^2\theta_{h,i}\right)\right].
\end{split}
\end{equation}

\begin{figure}[!htbp]
\begin{center}
\epsfig{file = 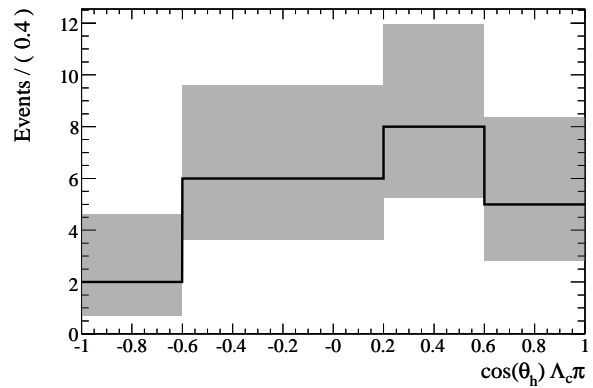, width=0.45\textwidth}
\renewcommand{\baselinestretch}{1}
\caption[Helicity angle distribution for the combined sample of background and non-resonant signal events.]{Helicity angle distribution for the combined sample of background and non-resonant signal events, a non-parameteric PDF (line). The shaded region indicates the Poisson uncertainty in the distribution.}
\label{fg:AngularBkg_HistPdf}
\end{center}
\end{figure}

The shape of the $\cos\theta_h$ distribution for background events is estimated from the shape of the helicity distribution for events in the background regions. The helicity distribution for these events is illustrated in Fig.~\ref{fg:AngularBkg_HistPdf} as a non-parametric PDF (a histogram). This PDF is used to generate the number of background events in the signal region with a Poisson uncertainty ($7.2 \pm 2.7$). The total number of events in the sample is fixed to $127$, so the background effectively dilutes the signal distribution.

\begin{figure}
\begin{center}
\epsfig{file = 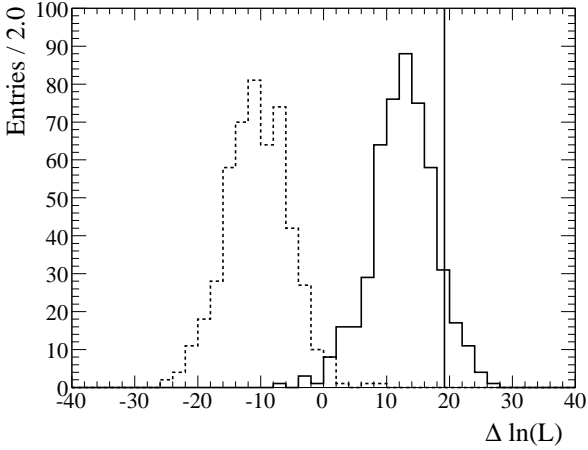, width=0.45\textwidth}
\renewcommand{\baselinestretch}{1}
\caption{Distribution of $\Delta\ln\calL = \ln\calL(1/2) - \ln\calL(3/2)$ for signal events generated with a uniform distribution in $\cos\theta_h$ (solid histogram, positive values) and a $1+3\cos^2\theta_h$ distribution (dashed histogram, negative values). Background events are included, and all events are efficiency-corrected. We measure $\Delta\ln\calL = +19.2$ in data (indicated by the vertical line), so we accept the spin-$1/2$ hypothesis.}
\label{fg:deltalogL_127f}
\end{center}
\end{figure}

We then generate 500 samples (127 events each) and compute the likelihood \calL that each generated distribution is uniform in $\cos\theta_h$ (spin-$1/2$) or distributed as $1+3\cos^2\theta_h$ (spin-$3/2$). We define the quantity $\Delta\ln\calL = \ln\calL(1/2) - \ln\calL(3/2)$. Figure~\ref{fg:deltalogL_127f} shows the distribution $\Delta\ln\calL$ for events generated with each hypothesis. The dashed histogram (negative values of $\Delta\ln\calL$) corresponds to samples generated according to the spin-$3/2$ hypothesis, and the solid histogram (positive values of $\Delta\ln\calL$) corresponds to samples generated according to the spin-$1/2$ hypothesis. For each distribution, the separation from zero illustrates how well we can discriminate between hypotheses given $127$ signal events.

\begin{figure}
\epsfig{file = 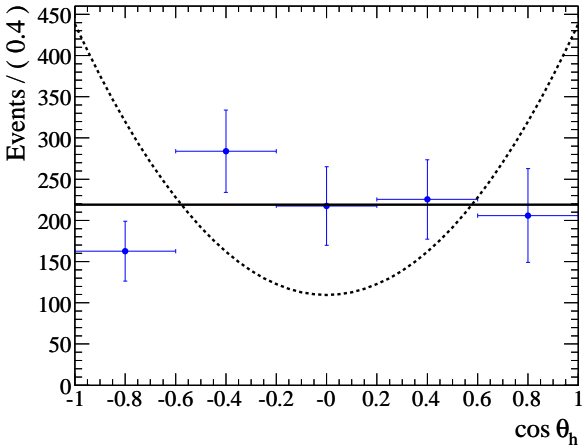, width=0.45\textwidth}
\renewcommand{\baselinestretch}{1}
\caption{The helicity angle distribution for \Sigmacz candidates in data. The points correspond to efficiency-corrected \BmtoSigcpb candidates. The curves for the spin-$1/2$ (solid line) and spin-$3/2$ (dashed line) hypotheses are overlaid.}
\label{fg:Sigc2455_angdist_data}
\end{figure}

The helicity angle distribution for events in the signal region around the \Sigmacz is shown in Figure~\ref{fg:Sigc2455_angdist_data}. The points are efficiency-corrected. Functions corresponding to the spin-$1/2$ (solid) and spin-$3/2$ (dashed) hypotheses are overlaid. We compute the difference in log likelihood between the hypotheses: $\Delta\ln\calL = +19.2$. We indicate the value of $\Delta\ln\calL$ in data with a vertical line in Figure~\ref{fg:deltalogL_127f}. The observed value of $\Delta\ln\calL$ is consistent with the spin-$1/2$ hypothesis and excludes the spin-$3/2$ hypothesis at the $>4\sigma$ level.

The ideal angular distributions for the \Sigmacz stated in Eqn.~\ref{eqn:AngDists} are also applicable for the excited \Sigcz resonance. Unlike the narrow \Sigmacz resonance near threshold, the excited \Sigcz is much wider and therefore its angular distribution is extremely contaminated by the non-resonant signal events underneath the signal. We perform a non-resonant sideband subtraction to extract the helicity angle distribution of the excited \Sigcz, but are limited by the number of signal events available. An examination of this distribution is somewhat consistent with a $J=1/2$ hypothesis, but no conclusive statement can be made about the spin of the observed excited \Sigcz.

\section{CONCLUSION}

We have presented branching fraction measurements for the decays \BzbtoLcpb and \BmtoLcpbpi:
\begin{multline}
\BR(\BzbtoLcpb) =\\
\shoveright{(1.89 \pm 0.21 \pm 0.06 \pm 0.49) \times 10^{-5},}\\
\shoveleft{\BR(\BmtoLcpbpi) =}\\
 (3.38 \pm 0.12 \pm 0.12 \pm 0.88) \times 10^{-4},
\end{multline}
where the uncertainties are statistical, systematic, and due to the uncertainty in $\BR(\LctopKpi)$, respectively. These measurements are based on $383$ million \BB\ events produced by the SLAC \BF\ and recorded by the \babar\ detector. 

If we combine the statistical and systematic uncertainties only, we obtain $\BR(\BzbtoLcpb) = (1.9\pm0.2)\times10^{-5}$, which is consistent with a previous measurement by the Belle Collaboration of $\BR(\BzbtoLcpb) = (2.2\pm 0.6)\times10^{-5}$~\cite{ref:belle2003}. Both measurements use the same value for $\BR(\LctopKpi)$. However, our measurement for the three-body mode, $\BR(\BmtoLcpbpi) = (3.4\pm0.2)\times10^{-4}$, is significantly larger (by about $4\sigma$) than the previous measurement from Belle $\BR(\BmtoLcpbpi) = (2.1\pm0.3)\times10^{-4}$~\cite{ref:belle2004}. The Belle Collaboration measurement uses six coarse regions across the \BmtoLcpbpi Dalitz plane to correct for variations in efficiency; we use much finer regions and see significant variation near the edges of the Dalitz plane. This difference in efficiency treatment may account for some of the discrepancy between the two results.

One of the main motivations for studying baryonic \B-meson decays is to gain knowledge about baryon-antibaryon production in meson decays. We have measured the ratio of the two branching fractions,
\begin{equation}
\frac{\BR(\BmtoLcpbpi)}{\BR(\BzbtoLcpb)} = 15.4 \pm 1.8 \pm 0.3.
\end{equation}
In this quantity the $26\%$ uncertainty in $\BR(\LctopKpi)$ cancels in the branching ratio. 

We have also measured the fractions of \BmtoLcpbpi decays that proceed through a \Sigmac resonance:
\begin{equation}
\begin{split}
\frac{\BR(\BmtoSigcpb)}{\BR(\BmtoLcpbpi)}         &= (12.3 \pm 1.2 \pm 0.8) \times 10^{-2}, \\
\frac{\BR(\BmtoSigcstarstarpb)}{\BR(\BmtoLcpbpi)} &= (11.7 \pm 2.3 \pm 2.4) \times 10^{-2}.
\end{split}
\end{equation}
Assuming no interference with direct decay to $\Lambdac\antiproton\pim$, about $1/4$ of \BmtoLcpbpi decays proceed through a \Sigmac resonance.

The order of magnitude difference between the decay rates of \BmtoLcpbpi and two-body decays such as \BzbtoLcpb, \BmtoSigcpb, and \BmtoSigcstarstarpb is consistent with the theoretical description~\cite{ref:housoni} that baryonic \B decays are favored when the baryon and antibaryon are close together in phase space. This interpretation is also supported by the observation of the enhancement in rate when \mLcp is near threshold. Although the \Lcp threshold enhancement alone could indicate a resonance below threshold, enhancements have been observed in other baryon-antibaryon systems and in decays such as $\epem \to \proton \antiproton \gamma$~\cite{ref:babarBtoppbarK,ref:babarBtoDppbar}. Therefore the body of measurements indicates that we are observing a phenomenon that is common to baryon production from meson decays, and possibly common to baryon production in general.

We have used the angular distribution of the decay \BmtoSigcpb to study the spin of the \Sigcz baryon. The helicity angle distribution is consistent with being uniform, which indicates that the \Sigcz has $J=1/2$ assuming the ground state \Lambdac also has $J=1/2$ and excludes the $J=3/2$ hypothesis at the $>4\sigma$ level. This is consistent with quark model expectations for the lowest \Sigmac baryon state.

We also observe an excited \Sigmac state in \B-meson decays to $\Lambdac\antiproton\pim$. We measure the mass of this resonance to be $(2846\pm8\pm10)\mevcc$ and the width to be $(86^{~+~33}_{~-~22})\mev$. It is possible that this observation is a confirmation of a triplet of $\Sigma_c(2800)$ states seen in \Lcpi continuum production~\cite{ref:BelleSigc2800}. However, the neutral \Sigmaczstarstar has a measured mass of $(2802^{~+~4}_{~-~7})\mevcc$ and width of $(61^{~+~28}_{~-~18})\mev$. The widths of the $\Sigma_{c}(2800)$ and the state observed in \B decays are consistent, but the masses are $3\sigma$ apart. If these are indeed the same state, then the discrepancy in measured masses needs to be resolved.

Another possible interpretation is that the excited \Sigcz resonance seen in this analysis is not the \Sigmaczstarstar that was previously observed. A clear signal is evident for \BmtoSigcpb decays, but we do not see any evidence for the decay \BmtoSigcstarpb. The absence of the decay \BmtoSigcstarpb is in contrast to a claimed $2.9\sigma$ signal from an analysis by the Belle Collaboration based on $152$ million \BB\ events~\cite{ref:belle2004}. Also, an examination of the \BmtoLcpbpi Dalitz plot shows no evidence for the decay \BmtoLambdaDelta. The \Sigmaczstar is a well-established state, and so is the $\Delta(1232)^{++}$. Both are expected to have $J=3/2$. The Belle Collaboration tentatively identified the \Sigmaczstarstar as $J=3/2$ based on the measured mass of the state, while there is weak evidence that the excited \Sigcz we observe is $J=1/2$. It is therefore possible that \B decays to higher-spin baryons are suppressed, perhaps due to the same baryon production mechanisms that suppress two-body baryonic decays, and that the excited \Sigcz state that we have observed is a newly-observed spin-$1/2$ state.

We are grateful for the 
extraordinary contributions of our \pep2\ colleagues in
achieving the excellent luminosity and machine conditions
that have made this work possible.
The success of this project also relies critically on the 
expertise and dedication of the computing organizations that 
support \babar.
The collaborating institutions wish to thank 
SLAC for its support and the kind hospitality extended to them. 
This work is supported by the
US Department of Energy
and National Science Foundation, the
Natural Sciences and Engineering Research Council (Canada),
the Commissariat \`a l'Energie Atomique and
Institut National de Physique Nucl\'eaire et de Physique des Particules
(France), the
Bundesministerium f\"ur Bildung und Forschung and
Deutsche Forschungsgemeinschaft
(Germany), the
Istituto Nazionale di Fisica Nucleare (Italy),
the Foundation for Fundamental Research on Matter (The Netherlands),
the Research Council of Norway, the
Ministry of Education and Science of the Russian Federation, 
Ministerio de Educaci\'on y Ciencia (Spain), and the
Science and Technology Facilities Council (United Kingdom).
Individuals have received support from 
the Marie-Curie IEF program (European Union) and
the A. P. Sloan Foundation.


\begin{thebibliography}{99}
\bibitem{ref:babarBtoppbarK}
B.\ Aubert {\em et al.} (\babar\ Collaboration), \jprd{72}, 051101 (2005).

\bibitem{ref:babarBtoDppbar}
B.\ Aubert {\em et al.} (\babar\ Collaboration), \jprd{74}, 051101 (2006).

\bibitem{ref:radBdecays_exp}
E.\ Barberio {\em et al.} (Heavy Flavor Averaging Group), \hepex{0603003} (2006); B.\ Aubert {\em et al.} ( \babar\ Collaboration), \jprl{97}, 171803 (2006); B.\ Aubert {\em et al.} (\babar\ Collaboration), \jprd{72}, 052004 (2005); P.\ Koppenburg {\em et al.} (Belle Collaboration), \jprl{93}, 061803 (2004); K.\ Abe {\em et al.} (Belle Collaboration), \plb{511}, 151 (2001); S.\ Chen {\em et al.} (CLEO Collaboration), \jprl{87}, 251807 (2001).

\bibitem{ref:belleBtoLpbargamma}
M.-Z.\ Wang {\em et al.} (Belle Collaboration), \jprd{76}, 052004 (2007).

\bibitem{ref:radBdecays_th}
M.\ Misiak {\em et al.}, \jprl{98}, 022002 (2007); 

T.\ Becher and M.\ Neubert, \jprl{98}, 022003 (2007).

\bibitem{ref:radBdecays_bary_th}
C.\ Q.\ Geng and Y.\ K.\ Hsiao, \plb{610}, 67 (2005); 

H.-Y.\ Cheng and K.-C.\ Yang, \plb{533}, 271 (2002).

\bibitem{ref:Argus1992}
H.\ Albrecht {\em et al.} (ARGUS Collaboration), \zpc{56}, 1 (1992).

\bibitem{ref:pdg2006}
W.-M.\ Yao {\em et al.} (Particle Data Group), \jpg{33}, 1 (2006) and 2007 partial update for the 2008 edition.

\bibitem{ref:Jarfipolemodel}
M.\ Jarfi {\em et al.}, \jprd{43}, 1599 (1991).

\bibitem{ref:Deshpandepolemodel}
N.\ G.\ Deshpande {\em et al.}, \mpl{A3}, 749 (1988).

\bibitem{ref:Balldiquarkmodel}
P.\ Ball and H.\ G.\ Dosch, \zpc{51}, 445 (1991).

\bibitem{ref:ReindersQCDsumrule}
L.\ J.\ Reinders {\em et al.}, \prep{127}, 1 (1985).

\bibitem{ref:CZQCDsumrule}
V.\ Chernyak and I.\ Zhitnitsky, \npb{345}, 137 (1990).

\bibitem{ref:chengpolemodel2003}
H.-Y.\ Cheng and K.-C.\ Yang, \jprd{67}, 034008 (2003).

\bibitem{ref:chengpolemodel2002}
H.-Y.\ Cheng and K.-C.\ Yang, \jprd{65}, 054028 (2002).

\bibitem{ref:housoni}
W.-S.\ Hou and A.\ Soni, \jprl{86}, 4247 (2001).

\bibitem{ref:ChargeConjImplied}
Unless specifically stated otherwise, conjugate decay modes are assumed throughout this paper.

\bibitem{ref:babarnim}
B.\ Aubert {\em et al.} (\babar\ Collaboration), \nima{479}, 1 (2002).

\bibitem{ref:EvtGen}
D.\ J. Lange, Nucl.\ Instrum.\ Methods Phys. Res., Sect. A {\bf 462}, 152 (2001).

\bibitem{ref:Jetset}
T. Sjostrand {\em et al.}, J. High Energy Phys. 05, (2006) 026.

\bibitem{ref:Geant4}
S. Agostinelli {\em et al.} (GEANT4 Collaboration), Nucl.\ Instrum.\ Methods Phys. Res., Sect. A {\bf 506}, 250 (2003).

\bibitem{ref:fisher}
R.\ A.\ Fisher, Annals Eugen.\ {\bf 7}, 179 (1936).

\bibitem{ref:thrust}
S.\ Brandt {\em et al.}, \jpl{12}, 57 (1964); E.\ Farhi, \jprl{39}, 1587 (1977).

\bibitem{ref:babarmmiss}
B.\ Aubert {\em et al.} (\babar\ Collaboration), \jprd{71}, 111102(R) (2005).

\bibitem{ref:belle2004}
N.\ Gabyshev {\em et al.} (Belle Collaboration), \jprl{97}, 242001 (2006).

\bibitem{ref:ARGUS}
H.\ Albrecht {\em et al.} (ARGUS Collaboration), \zpc{48}, 543 (1990).

\bibitem{ref:splot}
M. Pivk and F.\ R.\ Le Diberder, \nima{555}, 356 (2005).

\bibitem{ref:bluemethod}
L.\ Lyons {\em et al.}, \nima{270}, 110 (1988).

\bibitem{ref:BelleSigc2800}
R.\ Mizuk {\em et al.} (Belle Collaboration), \jprl{94}, 122002 (2005).

\bibitem{ref:belle2003}
N.\ Gabyshev {\em et al.} (Belle Collaboration), \jprl{90}, 121802 (2003).

\end{thebibliography}
\end{document}